\newif\ifplainstyle
\plainstyletrue
\plainstylefalse

\newif\ifjhepstyle
\jhepstyletrue
\newif\ifprstyle
\prstyletrue
\prstylefalse
\ifprstyle
	\documentclass[twocolumn,nofootinbib]{revtex4-1}
\else
	\documentclass[11pt,a4paper]{article}
\fi
\ifjhepstyle
	\usepackage{jheppub}
	\usepackage{amsfonts}
	\usepackage{verbatim}
	\usepackage{float}
	\usepackage{color}
	\renewcommand*\arraystretch{2}
	\usepackage{array}
	\newcolumntype{C}[1]{>{\centering\arraybackslash$}p{#1}<{$}}
	\makeatletter
	\def\@fpheader{\phantom{:-)}}
	\makeatother
\else	
 	\ifprstyle
		\usepackage{verbatim}
		\usepackage{amsmath,amsfonts,amssymb}
		\usepackage[colorlinks=true
                	,urlcolor=blue
                	,anchorcolor=blue
                	,citecolor=blue
                	,filecolor=blue
                	,linkcolor=blue
                	,menucolor=blue
                	]{hyperref}
	\else
            	\usepackage{verbatim}
            	\usepackage{cite}
            	\usepackage{setspace}
            	\usepackage[top=2.5cm, bottom=2.75cm, left=2.5cm, right=2.5cm]{geometry}
            	\usepackage{amsmath,amsfonts,amssymb}
            	\usepackage[colorlinks=true
            	,urlcolor=blue
            	,anchorcolor=blue
            	,citecolor=blue
            	,filecolor=blue
            	,linkcolor=blue
            	,menucolor=blue
            	,linktoc=page
            	]{hyperref}
            	\usepackage{float}
            	\restylefloat{table}
            	\renewcommand{\arraystretch}{1.5}
            	\numberwithin{equation}{section}
	\fi
\fi

\allowdisplaybreaks

\usepackage{tikz}
\usepackage{mathtools}
\usepackage{pgfplots}
\usepackage{subcaption}
\usepackage{bbold}
\usepackage{transparent}
\usepackage{bm}
\usepackage{enumitem}
\usepackage[textsize=footnotesize,textwidth=2cm]{todonotes}

\makeatletter
\let\save@mathaccent\mathaccent
\newcommand*\if@single[3]{%
  \setbox0\hbox{${\mathaccent"0362{#1}}^H$}%
  \setbox2\hbox{${\mathaccent"0362{\kern0pt#1}}^H$}%
  \ifdim\ht0=\ht2 #3\else #2\fi
  }
\newcommand*\rel@kern[1]{\kern#1\dimexpr\macc@kerna}
\newcommand*\widebar[1]{\@ifnextchar^{{\wide@bar{#1}{0}}}{\wide@bar{#1}{1}}}
\newcommand*\wide@bar[2]{\if@single{#1}{\wide@bar@{#1}{#2}{1}}{\wide@bar@{#1}{#2}{2}}}
\newcommand*\wide@bar@[3]{%
  \begingroup
  \def\mathaccent##1##2{%
    \let\mathaccent\save@mathaccent
    \if#32 \let\macc@nucleus\first@char \fi
    \setbox\z@\hbox{$\macc@style{\macc@nucleus}_{}$}%
    \setbox\tw@\hbox{$\macc@style{\macc@nucleus}{}_{}$}%
    \dimen@\wd\tw@
    \advance\dimen@-\wd\z@
    \divide\dimen@ 3
    \@tempdima\wd\tw@
    \advance\@tempdima-\scriptspace
    \divide\@tempdima 10
    \advance\dimen@-\@tempdima
    \ifdim\dimen@>\z@ \dimen@0pt\fi
    \rel@kern{0.6}\kern-\dimen@
    \if#31
      \overline{\rel@kern{-0.6}\kern\dimen@\macc@nucleus\rel@kern{0.4}\kern\dimen@}%
      \advance\dimen@0.4\dimexpr\macc@kerna
      \let\final@kern#2%
      \ifdim\dimen@<\z@ \let\final@kern1\fi
      \if\final@kern1 \kern-\dimen@\fi
    \else
      \overline{\rel@kern{-0.6}\kern\dimen@#1}%
    \fi
  }%
  \macc@depth\@ne
  \let\math@bgroup\@empty \let\math@egroup\macc@set@skewchar
  \mathsurround\z@ \frozen@everymath{\mathgroup\macc@group\relax}%
  \macc@set@skewchar\relax
  \let\mathaccentV\macc@nested@a
  \if#31
    \macc@nested@a\relax111{#1}%
  \else
    \def\gobble@till@marker##1\endmarker{}%
    \futurelet\first@char\gobble@till@marker#1\endmarker
    \ifcat\noexpand\first@char A\else
      \def\first@char{}%
    \fi
    \macc@nested@a\relax111{\first@char}%
  \fi
  \endgroup
}
\makeatother
\newcommand{\ThisIsTheTitle}{Heating up holography for single-trace $\bm{J\bar{T}}$ deformations} 
\newcommand{\ThisIsAuthorOne}{Luis Apolo}
\newcommand{\ThisIsEmailOne}{apolo@mail.tsinghua.edu.cn}
\newcommand{\ThisIsAuthorTwo}{and Wei Song}
\newcommand{\ThisIsEmailTwo}{wsong2014@mail.tsinghua.edu.cn}
\newcommand{\ThisIsTheAffiliation}{Yau Mathematical Sciences Center, Tsinghua University, Beijing 100084, China}
\newcommand{\TheseAreTheKeywords}{}
\newcommand{\ThisIsTheAbstract}{
We study thermodynamic aspects of a tractable toy model of holography for extremal Kerr black holes proposed in~\cite{Apolo:2018qpq}. On the gravity side, the theory can be described by the worldsheet action of string theory on a warped AdS$_3$ background supported by NS-NS flux. Once we turn on temperature, the deformed background is described by a black string solution of type IIB supergravity that features a locally warped AdS$_3$ factor. The dual field theory is conjectured to be a single-trace version of a $J\bar{T}$-deformed CFT at finite temperature. As evidence for the correspondence we show that the spectrum of strings winding on the deformed background agrees with the spectrum of $J\bar{T}$-deformed CFTs. Furthermore, we show that the gravitational charges of the black string match the averaged charges of a thermal ensemble in the dual field theory. Finally, we reproduce the Bekenstein-Hawking entropy of the black string from the microscopic density of states of $J\bar{T}$-deformed CFTs.}
\ifjhepstyle
\title{\ThisIsTheTitle}

\author{\ThisIsAuthorOne}
\author{\ThisIsAuthorTwo}
\affiliation{\ThisIsTheAffiliation}

\emailAdd{\ThisIsEmailOne}
\emailAdd{\ThisIsEmailTwo}
\abstract{\ThisIsTheAbstract} 

\fi
\begin{document}

%%%%%%%%%%%%%%%%%%%%%%%%%%%%%%%%%%%%%%%%%%%%%%
\ifjhepstyle
\maketitle
\flushbottom
%\raggedbottom
\fi
%%%%%%%%%%%%%%%%%%%%%%%%%%%%%%%%%%%%%%%%%%%%%%

%%%%%%%%%%%%%%%%%%%%%%%%%%%%%%%%%%%%%%%%%%%%%%
% Macros and definitions
%%%%%%%%%%%%%%%%%%%%%%%%%%%%%%%%%%%%%%%%%%%%%%
\long\def\symfootnote[#1]#2{\begingroup%
\def\thefootnote{\fnsymbol{footnote}}\footnote[#1]{#2}\endgroup} 

% Colored notes
\def\rednote#1{{\color{red} #1}}
\def\bluenote#1{{\color{blue} #1}}

% Definitions
\def\({\left (}
\def\){\right )}
\def\lb{\left [}
\def\rb{\right ]}
\def\lB{\left \{}
\def\rB{\right \}}

\def\Int#1#2{\int \textrm{d}^{#1} x \sqrt{|#2|}}
\def\Bra#1{\left\langle#1\right|} 
\def\Ket#1{\left|#1\right\rangle}
\def\BraKet#1#2{\left\langle#1|#2\right\rangle} 
\def\Vev#1{\left\langle#1\right\rangle}
\def\Vevm#1{\left\langle \Phi |#1| \Phi \right\rangle}\def\bbox{\bar{\Box}}
\def\til#1{\tilde{#1}}
\def\wtil#1{\widetilde{#1}}
\def\ph#1{\phantom{#1}}

\def\ra{\rightarrow}
\def\la{\leftarrow}
\def\lra{\leftrightarrow}
\def\p{\partial}
\def\barp{\bar{\partial}}
\def\diff{\mathrm{d}}

\def\sinh{\mathrm{sinh}}
\def\cosh{\mathrm{cosh}}
\def\tanh{\mathrm{tanh}}
\def\coth{\mathrm{coth}}
\def\sech{\mathrm{sech}}
\def\csch{\mathrm{csch}}

\def\a{\alpha}
\def\b{\beta}
\def\g{\gamma}
\def\d{\delta}
\def\e{\epsilon}
\def\ve{\varepsilon}
\def\k{\kappa}
\def\l{\lambda}
\def\n{\nabla}
\def\om{\omega}
\def\s{\sigma}
\def\t{\theta}
\def\z{\zeta}
\def\vp{\varphi}

\def\ss{\Sigma}
\def\dd{\Delta}
\def\GG{\Gamma}
\def\ll{\Lambda}
\def\tt{\Theta}

\def\A{{\cal A}}
\def\B{{\cal B}}
\def\C{{\cal C}}
\def\cE{{\cal E}}
\def\D{{\cal D}}
\def\F{{\cal F}}
\def\H{{\cal H}}
\def\I{{\cal I}}
\def\J{{\cal J}}
\def\K{{\cal K}}
\def\L{{\cal L}}
\def\N{{\cal N}}
\def\O{{\cal O}}
\def\Q{{\cal Q}}
\def\P{{\cal P}}
\def\cS{{\cal S}}
\def\W{{\cal W}}
\def\X{{\cal X}}
\def\Z{{\cal Z}}

\def\mfa{\mathfrak{a}}
\def\mfb{\mathfrak{b}}
\def\mfc{\mathfrak{c}}
\def\mfd{\mathfrak{d}}

\def\we{\wedge}
\def\re{\textrm{Re}}

\def\tilw{\tilde{w}}
\def\tile{\tilde{e}}

\def\tilL{\tilde{L}}
\def\tilJ{\tilde{J}}

\def\zz{\bar z}
\def\xx{\bar x}
\def\yy{\bar y}
\def\xp{x^{+}}
\def\xm{x^{-}}

\def\bp{\bar{\p}}
\def\note#1{{\color{red}#1}}
\def\notebf#1{{\bf\color{red}#1}}

\newcommand{\wei}[2]{\textcolor{red}{#1}\todo[color=cyan]{\scriptsize{#2}}}
\newcommand{\lui}[2]{\textcolor{red}{#1}\todo[color=orange]{\scriptsize{#2}}}

\def\VirU1{Vir \times U(1)}
\def\VirSL2R{\mathrm{Vir}\otimes\widehat{\mathrm{SL}}(2,\mathbb{R})}
\def\U1{U(1)}
\def\u1{U(1)}
\def\SL2R{\widehat{\mathrm{SL}}(2,\mathbb{R})}
\def\sl2r{\mathrm{SL}(2,\mathbb{R})}
\def\by{\mathrm{BY}}

\def\RR{\mathbb{R}}

\def\tr{\mathrm{Tr}}
\def\bnabla{\overline{\nabla}}

\def\sint{\int_{\ss}}
\def\dsint{\int_{\p\ss}}
\def\hint{\int_{H}}

% Equation environments
\newcommand{\eq}[1]{\begin{align}#1\end{align}}
\newcommand{\eqst}[1]{\begin{align*}#1\end{align*}}
\newcommand{\eqsp}[1]{\begin{equation}\begin{split}#1\end{split}\end{equation}}

% Square-root of determinants
\newcommand{\absq}[1]{{\textstyle\sqrt{\left |#1\right |}}}
%\newcommand{\absqeq}[1]{{\sqrt{|#1|}}}
%\newcommand{\absq}[1]{{\sqrt{-#1}}}

% Comment
%\newcommand{\comm}[1]{\begin{comment}#1\end{comment}}

%%%%%%%%%%%%%%%%%%%%%%%%%%%%%%%%%%%%%%%%%%%%%%

%%%%%%%%%%%%%%%%%%%%%%%%%%%%%%%%%%%%%%%%%%%%%%
% PR style titlepage
%%%%%%%%%%%%%%%%%%%%%%%%%%%%%%%%%%%%%%%%%%%%%%
\ifprstyle
\title{\ThisIsTheTitle}

\author{\ThisIsAuthorOne}
\email{\ThisIsEmailOne}

\author{\ThisIsAuthorTwo}
\email{\ThisIsEmailTwo}

\affiliation{\ThisIsTheAffiliation}

%\date{\today}

\begin{abstract}
\ThisIsTheAbstract
\end{abstract}

%\pacs{}

\maketitle

\fi
%%%%%%%%%%%%%%%%%%%%%%%%%%%%%%%%%%%%%%%%%%%%%%

%%%%%%%%%%%%%%%%%%%%%%%%%%%%%%%%%%%%%%%%%%%%%%
% Plain style titlepage and ToC
%%%%%%%%%%%%%%%%%%%%%%%%%%%%%%%%%%%%%%%%%%%%%%
\ifplainstyle
\begin{titlepage}
\begin{center}

\ph{.}

\vskip 4 cm

{\Large \bf \ThisIsTheTitle}

\vskip 1 cm

\renewcommand*{\thefootnote}{\fnsymbol{footnote}}

{{\ThisIsAuthorOne}\footnote{\ThisIsEmailOne} } and {\ThisIsAuthorTwo}\footnote{\ThisIsEmailTwo}

\renewcommand*{\thefootnote}{\arabic{footnote}}

\setcounter{footnote}{0}

\vskip .75 cm

%{\em \ThisIsTheAffiliation}

\end{center}

\vskip 1.25 cm
%\title{}
%\author{}
\date{}
%\maketitle

%\begin{abstract}
%\noindent \ThisIsTheAbstract
%\end{abstract}

\end{titlepage}

\newpage

\fi

\ifplainstyle
\tableofcontents
\noindent\hrulefill
\bigskip
\fi
%%%%%%%%%%%%%%%%%%%%%%%%%%%%%%%%%%%%%%%%%%%%%%

%%%%%%%%%%%%%%%%%%%%%%%%%%%%%%%%%%%%%%%%%%%%%%
% Draft titlepage
%%%%%%%%%%%%%%%%%%%%%%%%%%%%%%%%%%%%%%%%%%%%%%
\begin{comment}

%\begin{center}
%\Large{\textsf{\textbf{\ThisIsTheTitle}}}
%\end{center}

\begin{center}
{\Large \bf {\ThisIsTheTitle}}\\

\bigskip
\today

\end{center}

\vspace{-10pt}

\noindent\hrulefill

\noindent \ThisIsTheAbstract

\noindent\hrulefill

%\bigskip

\tableofcontents
\noindent\hrulefill
\bigskip

\end{comment}
%%%%%%%%%%%%%%%%%%%%%%%%%%%%%%%%%%%%%%%%%%%%%%

%%%%%%%%%%%%%%%%%%%%%%%%%%%%%%%%%%%%%%%%%%%%%%
\section{Introduction} \label{se:introduction}
%%%%%%%%%%%%%%%%%%%%%%%%%%%%%%%%%%%%%%%%%%%%%%

Our modern understanding of black holes is illuminated by holography, the profound observation that gravity is an emergent phenomenon described by a lower dimensional quantum field theory. In its most successful realization, the AdS/CFT correspondence~\cite{Maldacena:1997re,Gubser:1998bc,Witten:1998qj}, asymptotically anti-de Sitter (AdS) black holes are described by thermal states in a dual conformal field theory (CFT) \cite{Strominger:1996sh,Strominger:1997eq}. While anti-de Sitter space has played a fundamental role in the development of holography, the black holes found in nature are not asymptotically AdS, a fact that motivates extensions of holography beyond anti-de Sitter spacetimes.

One approach towards holography for realistic black holes is the Kerr/CFT correspondence~\cite{Guica:2008mu} (see~\cite{Bredberg:2011hp,Compere:2012jk} for reviews). Kerr/CFT conjectures the existence of a holographic description to extremal Kerr black holes, limiting solutions to the highly rotating black holes we expect to find in nature. A key feature of extremal Kerr are the $SL(2,R)_L \times \widebar{U(1)}_R$ isometries of its near horizon geometry, a property that is also shared by warped AdS$_3$, a spacetime of constant negative curvature that is not asymptotic to AdS$_3$. The study of holography for warped AdS$_3$~\cite{Anninos:2008fx,Compere:2009zj,Compere:2014bia,Song:2016gtd} and Kerr black holes~\cite{Guica:2008mu,Bredberg:2009pv,Castro:2010fd,Haco:2018ske,Haco:2019ggi} have lead to new insights into the workings of holography beyond AdS spacetimes. 

The Kerr/CFT correspondence deviates from AdS/CFT in other interesting ways. For instance, the dual field theory is not expected to be a local quantum field theory~\cite{Dias:2009ex,Guica:2010sw}. More concretely, we have learned from string theory constructions that the dual field theory can be described as a deformation of a CFT induced by an irrelevant operator with conformal dimension $(1,2)$~\cite{Guica:2010ej, Compere:2010uk, ElShowk:2011cm, Song:2011sr}. 

Recent developments in solvable irrelevant deformations of quantum field theories~\cite{Smirnov:2016lqw,Cavaglia:2016oda} shed new light into the Kerr/CFT correspondence. In particular, a simple and universal $(1,2)$ operator known as $J\bar{T}$ --- where $J$ is a chiral $U(1)$ current and $\bar{T}$ is a component of the stress tensor --- induces an integrable deformation in a CFT~\cite{Guica:2017lia}. A holographic duality between gravity in warped AdS$_3$ spacetimes and CFTs deformed by the single-trace version of $J\bar{T}$ has been proposed in~\cite{Apolo:2018qpq,Chakraborty:2018vja}, providing a tractable toy model of holography for extremal Kerr black holes.

The $J\bar{T}$ operator belongs to a class of irrelevant deformations that interpolate between two-dimensional CFTs in the IR and nonlocal but tractable QFTs in the UV~\cite{Smirnov:2016lqw,Cavaglia:2016oda,Guica:2017lia} (see~\cite{Jiang:2019hxb} for a review). 
Although $J\bar{T}$ is an irrelevant operator, the spectrum of $J\bar{T}$-deformed CFTs satisfies universal formulae that depend only on the undeformed spectrum, the deformation parameter, and the size of the cylinder where the theory is defined~\cite{Guica:2017lia,Chakraborty:2018vja}. Several universal features of $J\bar{T}$, as well as the Lorentz-preserving $T\bar{T}$ deformation~\cite{Smirnov:2016lqw,Cavaglia:2016oda}, have been studied in the literature including connections to 2d quantum gravity~\cite{Dubovsky:2012wk,Dubovsky:2013ira,Dubovsky:2017cnj,Cardy:2018sdv}, correlation functions~\cite{Giribet:2017imm,Kraus:2018xrn,Aharony:2018vux,Guica:2019vnb}, and partition functions~\cite{Dubovsky:2018bmo,Aharony:2018bad,Aharony:2018ics} among related studies and generalizations~\cite{Caselle:2013dra,Baggio:2018gct,Bonelli:2018kik,Chen:2018eqk,Cardy:2018jho,Conti:2018tca,Baggio:2018rpv,Chang:2018dge,Nakayama:2018ujt,Sun:2019ijq,Jiang:2019tcq,LeFloch:2019rut,Jiang:2019hux,Conti:2019dxg,Chakraborty:2019mdf,Nakayama:2019mvq,Frolov:2019nrr,Chang:2019kiu,Jeong:2019ylz,Coleman:2019dvf,LeFloch:2019wlf}. 

The holographic interpretation of $J\bar{T}$-deformed CFTs depends crucially on the nature of the deformation. In the context of the AdS/CFT correspondence, double-trace deformations do not change the AdS background but modify its boundary conditions, whereas single-trace deformations are capable of changing the local bulk geometry. This intuition is realized in both the double~\cite{Bzowski:2018pcy} and single-trace~\cite{Apolo:2018qpq,Chakraborty:2018vja} (see~\cite{Giveon:2019fgr} for a review) versions  of holography for $J\bar{T}$-deformed CFTs.\footnote{The holographic duals to $T\bar{T}$-deformed CFTs also depend on the double-trace~\cite{McGough:2016lol,Guica:2019nzm} or single-trace~\cite{Giveon:2017nie} character of the deformation. For the gravity modes, the negative sign of the double-trace deformation can be interpreted as imposing Dirichlet boundary conditions at finite radius in the bulk~\cite{McGough:2016lol}. Equivalently, both signs of the deformation can be interpreted as imposing mixed  boundary conditions at the asymptotic boundary~\cite{Guica:2019nzm}. As suggested in~\cite{Kraus:2018xrn,Hartman:2018tkw}, the cutoff interpretation in the presence of matter requires additional double-trace operators for the matter fields. 
%Under certain conditions, the double-trace deformation can be interpreted as introducing a finite cutoff in the bulk~\cite{McGough:2016lol} and more generally corresponds to a change of the boundary conditions~\cite{Guica:2019nzm}.
Holographic studies on $T\bar{T}$ deformations include~\cite{Giveon:2017myj,Shyam:2017znq,Asrat:2017tzd,Cottrell:2018skz,Chakraborty:2018kpr,Taylor:2018xcy,Donnelly:2018bef,Babaro:2018cmq,Hartman:2018tkw,Shyam:2018sro,Chakraborty:2018aji,Gorbenko:2018oov,Park:2018snf,Caputa:2019pam,Banerjee:2019ewu,Murdia:2019fax,Ota:2019yfe}.} 

The holographic dual to the single-trace $J\bar{T}$ deformation admits a weakly-coupled description as string theory on a warped AdS$_3 \times S^3$ background. The theory can be obtained from the NS-NS sector of type IIB string theory on AdS$_3 \times S^3$ by adding an exactly marginal deformation to the worldsheet action (related studies in this direction include \cite{Israel:2003ry,Israel:2004vv,Detournay:2005fz,Detournay:2010rh,Georgiou:2019jws}). Using the relationship between the vertex operators in string theory on AdS$_3$ and the conserved currents of the dual CFT~\cite{Kutasov:1999xu}, one can show that the marginal worldsheet deformation corresponds to a $J\bar{T}$ deformation of the dual theory. As evidence for the correspondence, one finds that the spectrum of strings winding on the warped AdS$_3 \times S^3$ background matches the spectrum of $J\bar{T}$-deformed CFTs~\cite{Apolo:2018qpq,Chakraborty:2018vja}.

In this paper, we strengthen the proposed duality between string theory on warped AdS$_3$ and single-trace $J\bar{T}$-deformed CFTs. This is achieved by turning on temperature in both the bulk and boundary sides of the correspondence, and showing that the string worldsheet and supergravity descriptions of the finite-temperature background are consistent with the spectrum and thermodynamics of $J\bar{T}$-deformed CFTs. 

The warped AdS$_3$ background studied in~\cite{Apolo:2018qpq} can be obtained from the massless BTZ black string via a TsT transformation, where \emph{T} stands for a T-duality and \emph{s} for a shift of coordinates. The finite-temperature generalization of this background can also be obtained  from generic BTZ black strings by a TsT transformation with the same parameter~\cite{Azeyanagi:2012zd}. Hence, both of these backgrounds are expected to belong to the same space of solutions in the deformed theory. In particular, we show that the warped BTZ background is equivalent to an instantaneous deformation of the worldsheet by the antisymmetric product of two Noether currents. This mirrors the definition of the $J\bar{T}$ deformation in the dual CFT which is also instantaneous and the antisymmetric product of two conserved currents. Furthermore, we will see that a closely related background --- one that is locally but not globally equivalent to the TsT solution --- can be interpreted as a marginal deformation of the worldsheet by the product of two chiral currents.

We study the warped BTZ black string from both its worldsheet and supergravity descriptions. In particular,
\vspace{-6pt}
\begin{itemize}
\item[(\emph{i})] we derive the spectrum of strings winding on the warped BTZ black string and recover the same worldsheet spectrum found at zero temperature. Consequently, we find that the string theory spectrum on the finite-temperature background matches the spectrum of $J\bar{T}$-deformed CFTs. Our results are sensitive to the global properties of the spacetime which we argue are determined by the TsT transformation.

\vspace{-6pt}

\item[(\emph{ii})] we compute the gravitational charges of the warped BTZ background and study its thermodynamics. We show that these charges match thermal expectation values in $J\bar{T}$-deformed CFTs and reproduce the Bekenstein-Hawking entropy of the black string from the density of states of the dual theory. A crucial step in these calculations is an appropriate identification of the deformed $U(1)$ charge which we determine using string theory. Our results show that the warped BTZ black string corresponds to a thermal state in the dual $J\bar{T}$-deformed CFT.
\end{itemize}
\vspace{-6pt}

The paper is organized as follows. In Section~\ref{se:fieldtheory} we study the thermodynamics of $J\bar{T}$-deformed CFTs and provide an alternative derivation of the deformed conformal weight. \linebreak[2] In Section~\ref{se:jtbar}, we review the duality between string theory on AdS$_3$ and $J\bar{T}$-deformed CFTs, and reproduce the conformal weight derived in the field theory analysis. In Section~\ref{se:warpedbtz} we argue that the warped BTZ black string obtained from a TsT transformation is equivalent to a single trace version of the $J\bar T$ deformation at finite temperature. In Section~\ref{se:worldsheet} we show that the spectrum of strings winding on the deformed background matches the spectrum of $J\bar{T}$ deformed CFTs. We consider the supergravity description of the warped BTZ black string in Section~\ref{se:entropy} where we compute its gravitational charges, thermodynamic potentials, and entropy. Therein we show that the warped BTZ background corresponds to a thermal state in the dual $J\bar{T}$-deformed CFT. In Section~\ref{se:globalads}, we list the results for the warped global AdS background, the latter of which can be obtained from analytic continuation of the results on warped BTZ. Finally, in Appendix~\ref{se:classicalstrings} we describe classical winding strings and check that their spectrum is consistent with the spectrum derived from the worldsheet; while in Appendix~\ref{se:currentcurrent} we show that after a shift of coordinates and a gauge transformation, the warped BTZ background is equivalent to an exactly marginal deformation of the worldsheet by the product of two chiral currents.

%%%%%%%%%%%%%%%%%%%%%%%%%%%%%%%%%%%%%%%%%%%%%%

%%%%%%%%%%%%%%%%%%%%%%%%%%%%%%%%%%%%%%%%%%%%%%
\section{The $J\bar{T}$ deformation} \label{se:fieldtheory}
%%%%%%%%%%%%%%%%%%%%%%%%%%%%%%%%%%%%%%%%%%%%%%

In this section we describe some of the salient features of the $J\bar{T}$ deformation of two-dimensional CFTs. In particular, we review the spectrum of the deformed theory, provide an alternative derivation of the deformed conformal weight, and study its thermodynamics. We will reproduce these features holographically in the following sections using the single trace version of the $J\bar{T}$ deformation in string theory.

%%%%%%%%%%%%%%%%%%%%%%%%%%%%%%%%%%%%%%%%%%%%%%
\subsection{Review of the spectrum} \label{suse:fieldtheoryspectrum}
%%%%%%%%%%%%%%%%%%%%%%%%%%%%%%%%%%%%%%%%%%%%%%

We begin by considering a two-dimensional CFT featuring a chiral $U(1)$ current $J \equiv J_x$, where $(x,\bar{x}) = (\vp + t, \vp - t)$  denote the left and right-moving coordinates of the CFT. The $J\bar{T}$ deformation is an irrelevant but solvable deformation of the theory defined by~\cite{Guica:2017lia}
  \eq{
  \frac{\p S_{CFT}}{\p \mu} = \int d^2 x \,{\cal O}_{J\bar{T}}({\mu}), \label{jtbardeformation}
  }
where $S_{CFT}$ denotes the action of the original CFT and $\mu$ is a parameter with dimensions of length. The $(1,2)$ operator ${\cal O}_{J\bar{T}}$ is evaluated instantaneously in the deformed theory and is defined by
  \eq{
  {\cal O}_{J\bar{T}} \equiv J \bar{T} - \bar{J} \tt, 
  \label{ojtbar}
  }
where $\bar{T} \equiv T_{\bar{x}\bar{x}}$ is the right-moving component of the stress tensor, $\tt \equiv T_{x\bar{x}}$, and $\bar{J} \equiv J_{\bar{x}}$ is the right-moving component of the $U(1)$ current, the latter of which is not necessarily chiral after the deformation. 

The deformation~\eqref{jtbardeformation} induces a flow from a conformal fixed point in the IR to a nonlocal field theory in the UV where $\mu$ parametrizes the scale of nonlocality. Like the $T\bar T$ operator~\cite{Zamolodchikov:2004ce}, the ${J\bar{T}}$ operator also factorizes~\cite{Guica:2017lia}, a feature that is crucial in the derivation of the deformed spectrum. In~\cite{Chakraborty:2018vja} it was assumed that (\emph{i}) the left-moving component of the stress tensor $T$ is chiral along the flow; (\emph{ii}) the $U(1)$ current $J$ also remains chiral after the deformation; and (\emph{iii}) the Sugawara combination of left-moving charges $T - J^2/4k$, where $k$ is the $U(1)$ level, is independent of $\mu$.\footnote{The $J\bar T $ operator can be understood as an antisymmetric product of two conserved currents, which leads to a derivation of the spectrum without the aforementioned assumptions~\cite{LeFloch:2019rut,LeFloch:2019wlf}. In particular, it is possible to define the $J\bar{T}$ operator using a $U(1)$ current that is not chiral along the flow. Nevertheless, when expressed in terms of the undeformed charges, the spectrum is equivalent to eqs.~\eqref{qft_spectrumleft} and~\eqref{qft_spectrumright2}.}  These assumptions, together with the fact that the angular momentum is unchanged by the deformation, lead to the following finite-size spectrum of the $J\bar{T}$-deformed theory~\cite{Guica:2017lia,Chakraborty:2018vja}
  \eq{
  %E_R(\mu) &= \frac{1}{2\mu^2 k} \bigg \{ \big[ R - \mu Q(0) \big] - \sqrt{ \big[ R - \mu Q(0) \big]^2 - 4 \mu^2 k R E_R(0)} \, \bigg \},   \label{qft_spectrumright}    \\
  E_L(\mu)-{Q(\mu)^2\over 4 R  k}&= E_L(0)-{Q(0)^2\over 4 R k}, \label{sugawara}\\
  E_L(\mu)-E_R (\mu) &=  E_L(0) - E_R(0)  , \label{qft_spectrumleft}\\
  Q(\mu)- 2\mu k  E_R(\mu)& = Q(0) ,  \label{qft_spectrumQ}
    }
where $E_L = \frac{1}{2} (E + P)$ and $E_R =  \frac{1}{2} (E - P)$ denote the left and right-moving energies ($E$ is the energy and $P$ is the momentum), $Q$ is the deformed $U(1)$ charge, and $2\pi R$ is the size of the cylinder.\footnote{Note that our left and right-moving energies differ by a factor of $2\pi$ from those used in~\cite{Guica:2017lia}, namely $E_{L,R}^{\,\textrm{here}} = (2\pi)^{-1} E_{L,R}^{\,\textrm{there}}$, while our $U(1)$ level differs by a factor of 4 such that $k^{\textrm{here}} = \frac{1}{4} k^{\textrm{there}}$. } 
The spectrum depends only on the deformation parameter $\mu$, the scale $R$, and the original CFT data $\{ E_L(0), E_R(0),Q(0)$\}; it is in this sense that the $J\bar{T}$ deformation is solvable. In particular, solving for $E_R(\mu)$ with the boundary condition $E_R(\mu)\big|_{\mu\to0}=E_R(0)$ yields
   \eq{E_R(\mu) &= \frac{1}{2\mu^2 k} \bigg \{ \big[ R - \mu Q(0) \big] - \sqrt{ \big[ R - \mu Q(0) \big]^2 - 4 \mu^2 k R E_R(0)} \, \bigg \}.   \label{qft_spectrumright2} 
   }
It is interesting to note that:   
\begin{enumerate}
\item[(\emph{i})] the spectrum becomes complex when the undeformed right-moving energy satisfies
  \eq{
  E_R(0) > \frac{\big[R - \mu Q(0) \big]^2}{4 \mu^2 k R}; \label{bound}
  }
\item[(\emph{ii})] when the undeformed $U(1)$ charge and angular momentum vanish, the deformed energy $E(\l)$ matches that of $T\bar{T}$-deformed CFTs with the negative sign of the deformation (in our conventions this is the sign that leads to superluminal propagation)~\cite{McGough:2016lol};
\item[(\emph{iii})] the energies and $U(1)$ charge of extremal solutions satisfying $E_R(0) = 0$ are left invariant by the deformation;
\item[(\emph{iv})] the torus partition function with the spectrum~\eqref{sugawara} --~\eqref{qft_spectrumQ} is modular covariant~\cite{Aharony:2018ics}. Conversely, assuming modular covariance of the partition function, one can derive the spectrum of $J\bar{T}$-deformed CFTs. 
\item[(\emph{v})]  if, for a fixed value of $\mu$, there exists a state in the undeformed theory satisfying $Q(0)=-\mu k E_R(0)$, then its left and right-moving energies are unchanged after the deformation, i.e.~$E_{L}(\mu)=E_{L}(0)$, $E_{R}(\mu)=E_{R}(0)$, and $Q(\mu) = - Q(0)$.
\end{enumerate}

%%%%%%%%%%%%%%%%%%%%%%%%%%%%%%%%%%%%%%%%%%%%%%

%%%%%%%%%%%%%%%%%%%%%%%%%%%%%%%%%%%%%%%%%%%%%%
\subsection{Conformal weight}
%%%%%%%%%%%%%%%%%%%%%%%%%%%%%%%%%%%%%%%%%%%%%%
The spectrum of $J\bar{T}$-deformed CFTs was derived on a cylinder of size $2 \pi R$. However, to calculate correlation functions and use conformal perturbation theory it is useful to work on the plane. The theory on the plane is specified by the left-moving conformal weight $h$, the $U(1)$ charge $Q$, and the right-moving energy $E_R$. In ref.~\cite{Guica:2019vnb}, the spectrum on the plane was obtained by taking an infinite boosted limit of the spectrum on the cylinder. We note that the $J\bar{T}$ deformation is expected to preserve an $SL(2,R)_L\times \widebar{U(1)}_R$ global symmetry and the full left-moving conformal symmetry of the original CFT.\footnote{The left conformal symmetry can be seen perturbatively in the field theory and holographically in both the double trace~\cite{Bzowski:2018pcy} and single trace~\cite{Azeyanagi:2012zd} (see Section~\ref{se:conformalweight}) versions of the $J\bar{T}$ deformation.} Here we would like to provide an alternative derivation of the conformal weight motivated by the study of warped conformal field theories~\cite{Hofman:2011zj,Detournay:2012pc} which share the same $SL(2,R)_L\times \widebar{U}(1)_R$ global symmetries of $J\bar T$-deformed CFTs. 

Conformal symmetry on the left-moving sector allows us to use an exponential map $x\to z=e^{ix/ R}$ to reformulate the $J\bar T$-deformed CFT in terms of the $(z,\bar x)$ coordinates. In this case, the cylinder identification $(x,\bar x)\sim (x+2\pi R, \bar x+2\pi R)$ is mapped to $(z, \bar x)\sim (e^{2\pi i}z, \bar x +2\pi R)$. Furthermore, the exponential map relates the left-moving conformal weight and energy in the usual way, namely
  \eq{
  h(\mu) \equiv R E_L(\mu)+{c\over24}. \label{cylinderplane}
  }
The formulae for the $J\bar{T}$-deformed spectrum in eqs.~\eqref{sugawara} --~\eqref{qft_spectrumQ} allows us to write the deformed conformal weight $h(\mu)$ in terms of the undeformed one $h(0)$ such that
  \eq{
  h(\mu) &= h(0) + {\mu} Q(0) E_R(\mu) + \mu^2 k E_R(\mu)^2, \label{defconformalweight}
  }
an expression that is valid for arbitrary values of the radius. In particular, we would like to take the plane limit $R\to\infty$ assuming that $\mu$, $Q(0)$, and $E_R(0)$ remain finite. In this limit the right-moving energy is independent of the deformation and given by
  \eq{
  \lim_{R\to\infty} {E_R(\mu)}=E_R(0).
  }
As $R\to \infty$, $\p_{\bar z} \to i R\p_{\bar x}$, where $\bar z=e^{-i \bar x / R}$ is the complex conjugate of $z$, and we can perform conformal perturbation on the complex plane $(z,\bar z)$ where operators are labelled by the conformal weight $h(0)$, the $U(1)$ charge $Q(0)$, and the right-moving energy $E_R(0)$. It follows that the deformed spectrum on the plane is given by 
  \eq{
   h(\mu) &= h(0) + {\mu} Q(0) E_R(0)+ \mu^2 k E_R(0)^2, \label{defconformalweight2} \\ 
   Q(\mu) &=Q(0)+2\mu k E_R(0),
  }
which agrees with the derivation of ref.~\cite{Guica:2019vnb}. In Section~\ref{se:jtbar}, we will provide a holographic justification for the relationship between the deformed left-moving energy $E_L(\mu)$ and the deformed conformal weight $h(\mu)$ in eq.~\eqref{cylinderplane}. More concretely, we will show that $E_L$ is the eigenvalue of the zero mode generator $L_0$ of the left-moving conformal symmetry preserved by the deformation.

%%%%%%%%%%%%%%%%%%%%%%%%%%%%%%%%%%%%%%%%%%%%%% 

%%%%%%%%%%%%%%%%%%%%%%%%%%%%%%%%%%%%%%%%%%%%%% 
\subsection{Thermodynamics} \label{suse:thermodynamics}
%%%%%%%%%%%%%%%%%%%%%%%%%%%%%%%%%%%%%%%%%%%%%%

The density of states of $J\bar{T}$-deformed CFTs is not expected to change under the deformation. This follows from the fact that there is no level crossing in the formulae for the deformed spectrum. In particular, the asymptotic density of states in the undeformed theory is given by the Cardy formula which in our conventions reads
  \eq{
  S_{CFT} = 2 \pi  \Bigg\{ \sqrt{\frac{c}{6} \bigg [R E_L(0) -\frac{Q(0)^2}{4k} \bigg]} + \sqrt{ \vphantom{\bigg]} \frac{c}{6} R E_R(0)}  \, \Bigg \} =   \frac{\pi^2 R c}{3}  \big[ T_L(0) + T_R(0) \big], \label{cardy}
  }
where $T_{L,R}(0)$ are the temperatures conjugate to the undeformed energies $E_{L,R}(0)$ defined via the thermodynamic relations
  \eq{
R E_L(0) - \frac{Q(0)^2}{4k} = \frac{c}{6} \big[ \pi R T_L(0)\big]^2, \qquad R E_R(0) = \frac{c}{6} \big[ \pi R T_R(0)\big]^2. \label{cftthermo}
  }
Assuming that the entropy does not change after the deformation, the spectrum of $J\bar{T}$-deformed CFTs allows us to write the entropy in the microcanonical ensemble such that\footnote{There is a subtlety in the definition of asymptotic states in the deformed theory since the energies become imaginary for large $E_R(0)$. Nevertheless, a precise definition is available in terms of the undeformed energies for which the entropy~\eqref{cardy} remains well defined for arbitrarily large $E_R(0)$.}
  \eq{
  \!\!\!  \!\!S_{J\bar{T}} = & S_{CFT} \notag \\
  =  & 2 \pi  \Bigg\{ \! \sqrt{\frac{c}{6} \bigg[ R E_L(\mu) - \frac{Q(\mu)^2}{4 k} \bigg] \!} + \sqrt{\frac{c}{6} \bigg[ R E_R(\mu) + {\mu E_R(\mu)} \big[\mu k E_R(\mu) - Q(\mu)\big] \bigg] \!}  \Bigg\}.  \label{deformedentropy}
  }
The contribution from the left-moving sector to the entropy takes the same form as in the undeformed theory. This is a consequence of the fact that the Sugawara combination of left-moving charges is independent of the deformation. Generically, the spectrum of $J\bar{T}$-deformed CFTs features a nonzero $U(1)$ charge $Q(\mu)$ even when the undeformed $U(1)$ charge vanishes, see eq.~\eqref{qft_spectrumleft}. Hence, we expect the first law of thermodynamics to satisfy
  \eq{
  \d S_{J\bar{T}} = \frac{1}{T_L(\mu)} \d E_L(\mu) + \frac{1}{T_R(\mu)} \d E_R(\mu) + \nu(\mu) \d Q(\mu),
  }
where $\nu(\mu)$ is the chemical potential conjugate to $Q(\mu)$. The thermodynamic potentials of the deformed theory must therefore satisfy
%%%%%%%%%%%%%%%%%%%%%%%%%%%%%%%%%%%%%%%%%%%%%%
%%%%%%%%%%%%%%%%%%%%%%%%%%%%%%%%%%%%%%%%%%%%%%
%%%%%%%%%%%%%%%%%%%%%%%%%%%%%%%%%%%%%%%%%%%%%%
\begin{comment}
%%%%%%%%%%%%%%%%%%%%%%%%%%%%%%%%%%%%%%%%%%%%%%
\footnote{Note that it is possible to express $\nu(\mu)$ entirely in terms of the undeformed thermodynamic potentials but the resulting expression is not particularly illuminating. Nevertheless, for completeness we have
  %
    \eqst{
  \nu (\mu) = \nu (0) - \frac{1}{2\mu k} \bigg[ \frac{1}{T_L(0)} + \frac{1}{T_R(0)} \bigg]  \bigg \{ 1 + 2\mu k T_L(0) \nu(0) - \sqrt{ \big[ 1 + 2\mu k T_L(0) \nu(0) \big]^2 -4 \mu^2 k  \frac{c}{6} \pi^2 T_R(0)^2 } \bigg\}.
 }
 %  
 }
 %%%%%%%%%%%%%%%%%%%%%%%%%%%%%%%%%%%%%%%%%%%%%%
\end{comment}
%%%%%%%%%%%%%%%%%%%%%%%%%%%%%%%%%%%%%%%%%%%%%%
%%%%%%%%%%%%%%%%%%%%%%%%%%%%%%%%%%%%%%%%%%%%%%
%%%%%%%%%%%%%%%%%%%%%%%%%%%%%%%%%%%%%%%%%%%%%%
  %
  \eq{
  \frac{1}{T_L(\mu)} &= \frac{\p S_{J\bar{T}}}{\p E_L(\mu)} = \frac{\pi R}{\sqrt{\frac{6}{c} \Big[ R E_L(0) - \tfrac{Q(0)^2}{4k} \Big]}} = \frac{1}{T_L(0)}, \label{deformedTL}\\
   \frac{1}{T_R(\mu)} &= \frac{\p S_{J\bar{T}}}{\p E_R(\mu)} = \frac{\pi R}{\sqrt{\frac{6}{c} R E_R(0)}} \Big[1 - \frac{\mu}{R}Q(0) \Big] = \frac{1}{T_R(0)}  \big[1 - \frac{\mu}{R}Q(0) \big] , \label{deformedTR}\\
{\nu(\mu)}&= \frac{\p S_{J\bar{T}}}{\p Q(\mu)} = -\frac{1}{T_L(0)} \frac{Q(\mu)}{2k R} - \frac{1}{T_R(0)} \frac{Q(\mu)-Q(0)}{2k R},  \label{deformednu}
  }
where we note that it is possible to write $\nu(\mu)$ in terms of the undeformed thermodynamic potentials but the resulting expression is not particularly illuminating.
Also note that the left-moving temperature is unchanged after the deformation, which is again a consequence of the $\mu$-independence of the Sugawara combination of left-moving charges. 

%%%%%%%%%%%%%%%%%%%%%%%%%%%%%%%%%%%%%%%%%%%%%%
%%%%%%%%%%%%%%%%%%%%%%%%%%%%%%%%%%%%%%%%%%%%%%
%%%%%%%%%%%%%%%%%%%%%%%%%%%%%%%%%%%%%%%%%%%%%%
\begin{comment}
In particular, when the undeformed $U(1)$ charge vanishes, the thermodynamic potentials simplify and are given by  
  %
  \eq{
  T_L(\mu) \big|_{Q(0) = 0} &= T_L(0), \qquad \qquad  T_R(\mu)\big|_{Q(0) = 0} = T_R(0),  \label{deftemperatures} \\
  \nu(\mu)\big|_{Q(0) = 0} &= -\bigg(\frac{1}{T_L(0)} + \frac{1}{T_R(0)} \bigg) \frac{Q(\mu)}{2 k R}.
  }
  %
\end{comment}
%%%%%%%%%%%%%%%%%%%%%%%%%%%%%%%%%%%%%%%%%%%%%%
%%%%%%%%%%%%%%%%%%%%%%%%%%%%%%%%%%%%%%%%%%%%%%
%%%%%%%%%%%%%%%%%%%%%%%%%%%%%%%%%%%%%%%%%%%%%%

While it is possible to express the entropy in the canonical ensemble, i.e.~entirely in terms of the deformed temperatures and chemical potential, for later convenience we write the entropy in a mixed ensemble with fixed temperatures $T_{L,R}(\mu)$ and $U(1)$ charge $Q(\mu)$. In this ensemble the entropy satisfies
 \eq{
 {S}_{J\bar{T}} &= \frac{\pi^2 R c}{3}  \bigg \{ T_L(\mu) +  \g\, T_R(\mu)  \Big[1 - \frac{\mu}{R}Q(\mu) \Big] \bigg \}, \label{entropyfieldtheory}
  %&=   \frac{\pi^2 R c}{3}  \Big \{ T_L(\mu) + T_R(\mu)  \Big[1 - \frac{\mu}{R}Q(0) \Big]\Big \}, 
 }
where $\g$ is defined by
  \eq{
  \frac{1}{\g} = \sqrt{1 - \frac{2 \pi^2 c\, k}{3} \mu^2 T_R(\mu)^2}.
  }
  %
%%%%%%%%%%%%%%%%%%%%%%%%%%%%%%%%%%%%%%%%%%%%%%
%%%%%%%%%%%%%%%%%%%%%%%%%%%%%%%%%%%%%%%%%%%%%%
%%%%%%%%%%%%%%%%%%%%%%%%%%%%%%%%%%%%%%%%%%%%%%
\begin{comment}
Interestingly, when the undeformed $U(1)$ charge vanishes the entropy is unchanged by the deformation and given by the Cardy formula~\eqref{cardy}, as can be seen by writing $\til{S}_{J\bar{T}}$ as
  %
  \eq{
   \til{S}_{J\bar{T}} & =   \frac{\pi^2 R c}{3}  \Big \{ T_L(\mu) + T_R(\mu)  \Big[1 - \frac{\mu}{R}Q(0) \Big]\Big \}. \label{entropyfieldtheory2}
  }
  % 
\end{comment}
%%%%%%%%%%%%%%%%%%%%%%%%%%%%%%%%%%%%%%%%%%%%%%
%%%%%%%%%%%%%%%%%%%%%%%%%%%%%%%%%%%%%%%%%%%%%%
%%%%%%%%%%%%%%%%%%%%%%%%%%%%%%%%%%%%%%%%%%%%%%
We note that when $T_R(\mu) = 0$, the entropy takes the same form in both the deformed and undeformed theories. This is consistent with the fact that the spectrum of extremal states satisfying $E_R(0) = 0$ is unchanged by the deformation. Finally, in the mixed ensemble, the expectation values of the conjugate energies $E_{L,R}(\mu)$ and the chemical potential $\nu (\mu)$ are given in terms of $T_{L,R}(\mu)$ and $Q(\mu)$ by
  \eq{
  E_L(\mu) &= \frac{Q(\mu)^2}{4k R} + \frac{\pi^2 R c}{6} T_L(\mu)^2, \label{averageEL} \\
  E_R(\mu) &= \frac{R}{2 \mu^2 k} \big( \g - 1 \big) \bigg[ 1 - \frac{\mu}{R} Q(\mu) \bigg], \label{averageER} \\
  \nu(\mu) &= -\frac{1}{T_L(\mu)} \frac{Q(\mu)}{2k R} - \frac{1}{T_R(\mu)} \frac{1}{2 \mu k} \bigg( 1 - \frac{1}{\g} \bigg). \label{averagenu}
  }
  %

%%%%%%%%%%%%%%%%%%%%%%%%%%%%%%%%%%%%%%%%%%%%%% 

%%%%%%%%%%%%%%%%%%%%%%%%%%%%%%%%%%%%%%%%%%%%%%
\section{String theory and $J\bar{T}$ deformations} \label{se:jtbar}
%%%%%%%%%%%%%%%%%%%%%%%%%%%%%%%%%%%%%%%%%%%%%%

In this section we consider a marginal deformation of string theory on AdS$_3\times S^3 \times {\cal M}^4$ that corresponds to the single trace $J\bar{T}$ deformation of the dual CFT~\cite{Apolo:2018qpq,Chakraborty:2018vja}. We begin by reviewing and adapting the results of ref.~\cite{Apolo:2018qpq} to our conventions. We then derive the deformed conformal weight of the dual $J\bar{T}$-deformed theory directly from the worldsheet. Readers familiar with the story at zero temperature can skip Section~\ref{se:jtbarspectrum}.

%%%%%%%%%%%%%%%%%%%%%%%%%%%%%%%%%%%%%%%%%%%%%%
\subsection{The warped massless BTZ background} \label{se:jtbarspectrum}
%%%%%%%%%%%%%%%%%%%%%%%%%%%%%%%%%%%%%%%%%%%%%%

Our starting point is the bosonic sector of string theory on AdS$_3\times S^3\times {\cal M}^4$ backgrounds supported by NS-NS flux. The compact four-dimensional manifold ${\cal M}^4$ does not play a role in what follows so we focus on the six-dimensional part of the background. In the Einstein frame, the action of type IIB supergravity is given by
  \eq{
  S_{\textrm{IIB}} = \frac{1}{16 \pi G_6} \int d^6x \sqrt{|g|} \bigg( R - \p_{\mu} \Phi \p^{\mu} \Phi - \frac{e^{-2 \Phi}}{12} H_{\mu\nu\a}H^{\mu\nu\a} \bigg ), \label{sugra}
  }
where $H$ is the field strength of the Kalb-Ramond $B$ field, $\Phi$ is the dilaton, and the Einstein ($g_{\mu\nu}$) and string frame ($G_{\mu\nu}$) metrics are related by
  \eq{
  g_{\mu\nu} = e^{-{\Phi}} G_{\mu\nu}. \label{einsteinstringframe}
  }
In our conventions the background metric, $B$-field, and dilaton are given by
  \eq{
 {ds^2}{} &=  \ell^2 \bigg \{ {dr^2\over 4 r^2} + r du dv+ \frac{1}{4} \Big [  \( d \chi + \cos\t \, d \phi \)^2 + d\t^2 + \sin^2 \t \, d \phi^2 \Big ]  \bigg \}, \label{mbtzmetric} \\
  B &= {\ell^2 \over4} \Big ( \cos \t\, d\phi \we d \chi - 2r du \we dv \Big ), \label{mbtzbfield} \\
  e^{-2\Phi} &= 1, \label{mbtzdilaton}
  }
where $\ell$ is the AdS scale, $(u = \vp + t/\ell, \,v = \vp - t/\ell, \, r)$ are the coordinates of the massless BTZ black hole satisfying
  \eq{
  (u,v) \sim (u + 2 \pi, v + 2\pi), \label{mbtzspatialcircle}
  }
and $(\chi, \phi, \t)$ parametrize the $S^3$ as a Hopf fibration over an $S^2$ with
  \eq{
  \chi \sim \chi + 4 \pi, \qquad \phi \sim \phi + 2 \pi, \qquad \t \sim \t + \pi.
  }

The bosonic part of the worldsheet action is given by
  \eq{
  S_0 = k \int dz^2 \bigg\{ \frac{\p r \bp r}{4 r^2} + r \p v \bp u + \frac{1}{4} \Big [ \big(\p\chi + 2 \cos\t\,\p\phi \big) \bp \chi  +  \p\t\bp\t + \p\phi\bp\phi  \Big] \bigg \}, \label{mbtzaction}
  }
where $\ell_s = \ell / \sqrt{k}$ is the string scale, $(z = \tau + \s, \bar{z} = \tau - \s)$ are the worldsheet coordinates, and $\p \equiv \p_z$, $\bp \equiv \p_{\bar{z}}$. The action features an affine ${SL}(2,R)_L \times SU(2)_L \times \widebar{SL(2,R)}_R \times \widebar{SU(2)}_R$ symmetry group whose global part matches  the isometries of the background. In particular, the $SL(2,R)$ currents $j^-$ and $\bar{j}^-$ coincide with the Noether currents generating translations along $u$ and $v$, and are given by
  \eq{
  j^{-}(z) = k r \p v, \qquad \bar{j}^{-}(\bar{z}) = k r \bp u, \label{jminus}
  }
while the $SU(2)$ currents $k^3$ and $\bar{k}^3$, which are related to translations along $\chi$ and $\phi$, read
  \eq{
  k^3 (z) = \frac{k}{2} \big( \p \chi + \cos\t\, \p\phi \big), \qquad \bar{k}^3 (\bar{z}) = \frac{k}{2} \big( \bp \phi + \cos\t\, \bp \chi \big). \label{k30}
  }
Note that the $k^3$ and $\bar{k}^3$ currents are spacelike while the $j^-$ and $\bar{j}^-$ currents are null. In Section~\ref{se:warpedbtz} we will consider BTZ backgrounds with nonvanishing mass and angular momentum where the currents generating translations along $u$ and $v$ become spacelike. In contrast, for the global AdS vacuum considered in Section~\ref{se:globalads} these currents are timelike. 

There is evidence that the long string sector of string theory on AdS$_3 \times S^3 \times {\cal M}^4$ is dual to a symmetric product CFT~\cite{Argurio:2000tb,Giveon:2005mi,Eberhardt:2019qcl} (see also the discussion in~\cite{Chakraborty:2019mdf})
  \eq{
  \big ( {\cal M}_{6k} \big )^p/S_p,  \label{symprod}
  }
where the seed theory ${\cal M}_{6k}$, which depends on the internal manifold ${\cal M}^4$, has central charge $c = 6k$. As a result, the total central charge of the dual CFT is given by $c_{st} = 6 k p $ where $k$ and $p$ denote the magnetic and electric charges of the background. In particular, the worldsheet theory features vertex operators describing observables with scaling dimension $(1,2)$ in the dual CFT. These operators, which are irrelevant and break Lorentz invariance from the CFT point of view, are given by~\cite{Kutasov:1999xu}
 \eq{
  A^a (x,\bar{x}) = \frac{1}{2} \int d^2 z \Big [ \p_{{x}} \bar{J}(\bar{x};\bar{z}) \p_{\bar{x}} +2 \p_{\bar{x}}^2 \bar{J}(\bar{x};\bar{z}) \Big ] \Phi_1(x,\bar{x};z,\bar{z}) {k}^a({z}), \label{Aoperator}
  }
where $(x,\bar x)$ are the coordinates of the dual CFT, $\bar{J}(\bar{x};\bar{z})$ is an $\bar{x}$-dependent linear combination of $\widebar{SL(2,R)}_R$ worldsheet currents, ${k}^{a}({z})$ denotes an ${SU(2)}_L$ current, and $\Phi_1(x,\bar{x}; z,\bar{z})$ can be interpreted as a bulk-to-boundary propagator~\cite{Kutasov:1999xu}.\footnote{In more detail, $\bar{J}(\bar{x};\bar{z}) = 2 \bar{x} \bar{j}^3 (\bar{z}) -  \bar{j}^+ (\bar{z})  - \bar{x}^2  \bar{j}^- (\bar{z})$ where $\bar{j}^a (\bar{z})$ are $\widebar{SL(2,R)}_R$ currents.} The $(1,2)$ operator~\eqref{Aoperator} shares the same quantum numbers as the $J\bar{T}$ deformation proposed in~\cite{Guica:2017lia}, and it was shown to correspond to a single-trace $J\bar{T}$ deformation of the dual CFT in~\cite{Apolo:2018qpq,Chakraborty:2018vja}. The operator~\eqref{Aoperator} is local on the worldsheet and leads to a marginal current-current deformation
  \eq{
 \dd S_{\l} = -\frac{2\l}{k} \int d^2x A^3(x,\bar{x}) =  \frac{2 \l}{ k} \int d^2 z \, {{k}}^3({z}) \bar{j}^{-}(\bar{z}),\label{jtbar}
  }
where $\l$ is a dimensionless constant. The relationship between the worldsheet and field theory deformation parameters $\l$ and $\mu$ is given by
  \eq{
  \ell \l = \mu k. \label{mulambda}
  }
Not surprisingly, the worldsheet currents featured in eq.~\eqref{jtbar} are related to translations along $\chi$ and $v$, reflecting the nature of the $J$ and $\bar{T}$ components of the $J\bar{T}$ operator.

The deformed worldsheet action is given by $S_{\l} = S_0 + \dd S_{\l}$ and describes strings on the warped massless BTZ$\times S^3$ background
  \eq{
 {ds^2}{} &=  \ell^2 \bigg \{ {dr^2\over 4 r^2} + r du \Big[dv + {\l} \big( d\chi + \cos\t\,d \phi \big) \Big] + d \Omega_3^2  \bigg \}, \label{mbtzmetricdef} \\
  B & = {\ell^2 \over4} \bigg \{ \cos \t\, d\phi \we d \chi - 2r du \we \Big[ dv +  {\l} \big( d \chi + \cos\t\,d\phi \big) \Big ] \bigg \}, \label{mbtzbfielddef}\\
 e^{-2 \Phi}&=1,
  }
where $d \Omega_3^2$ denotes the metric of the $S^3$. Since $\bar{j}^-$ and $k^3$ belong to commuting $\widebar{SL(2,R)}_R$ and $SU(2)_L$ algebras, the deformation is exactly marginal~\cite{Chaudhuri:1988qb} and eqs.~\eqref{mbtzmetricdef} and~\eqref{mbtzbfielddef} describe exact string theory backgrounds. The deformed metric~\eqref{mbtzmetricdef} admits only an $SL(2,R)_L \times U(1)_L \times \widebar{U(1)}_R \times \widebar{SU(2)}_R$ subgroup of the original $SL(2,R)_L \times SU(2)_L \times \widebar{SL(2,R)}_R \times \widebar{SU(2)}_R$ isometries and its connection to warped AdS$_3$ and the Kerr/CFT correspondence is discussed in detail in~\cite{Bena:2012wc,Azeyanagi:2012zd,Apolo:2018qpq}. 

Interestingly, the deformed background fields~\eqref{mbtzmetricdef} and~\eqref{mbtzbfielddef} can also be obtained from the original ones via the solution-generating TsT transformation of~\cite{Lunin:2005jy,Maldacena:2008wh}. The TsT transformation consists of T-duality along the $\chi$ coordinate, a shift $v' \to t^- + \frac{2\l}{k} \psi$ and $\chi' \to \psi$ where $(v', \chi')$ denote the T-dual coordinates, and another T-duality along $\psi$. The $t^-$ and $\psi$ coordinates characterize the TsT background and have been introduced here for future convenience. In order to write the TsT-transformed solution in the form of eqs.~\eqref{mbtzmetricdef} and~\eqref{mbtzbfielddef}, an additional change of coordinates is necessary such that
  \eq{
  t^- = v + \frac{\l}{2} \chi, \qquad \psi = \chi. \label{shift2}
  }
We will refer to the whole procedure, namely a TsT transformation followed by the additional shift of coordinates~\eqref{shift2}, as a \emph{TsTs} transformation.\footnote{The relationship between TsT transformations and marginal deformations has been explored further in~\cite{Araujo:2018rho}. Therein, it is shown that the additional shift~\eqref{shift2} that is necessary to reproduce the marginal deformation~\eqref{jtbar} can be obtained from a generalization of $O(d,d)$ transformations.} The TsTs transformation is locally equivalent to the current-current deformation of the worldsheet~\eqref{jtbar}. However, a subtlety arises in the identification of the deformed background due to the change of coordinates~\eqref{shift2}, and one may wonder whether the TsT background or the TsTs solution corresponds to the $J\bar T$ deformation of the dual field theory. At zero temperature, the analysis of~\cite{Apolo:2018qpq,Chakraborty:2018vja} cannot differentiate between the two backgrounds, and hence either proposal works. However, as we will show later, at finite temperature the difference plays an important role and it is the TsT background that is dual to a thermal state in the dual $J\bar T$-deformed CFT.

The spectrum of the deformed theory can be derived using spectral flow, as described in detail in Section~\ref{se:worldsheet}, and is found to satisfy~\cite{Apolo:2018qpq,Chakraborty:2018vja}
  \eq{
  E_R(0) &= E_R(\l) + \frac{\l E_R(\l) }{w k} \big [ { \l \ell  } E_R(\l) + Q(0) \big], \label{ogspectrumleft}\\
  E_L(0) &= E_L(\l) + \frac{ \l  E_R(\l) }{w k} \big [ \l \ell   E_R(\l) + Q(0) \big], \label{ogspectrumright}\\
  Q(0) &= Q(\l) - 2 \l  \ell  E_R(\l), \label{ogspectrumu1}
    }
where $E_{L}$ and $E_{R}$ are the left and right-moving energies, $Q$ is the $U(1)$ charge, $4k$ is the $U(1)$ level, and $2\pi \ell$ is the size of the cylinder at the boundary. The parameter $w$ in eqs.~\eqref{ogspectrumleft} and~\eqref{ogspectrumright} is an integer whose absolute value can be interpreted as the number of times the string winds along the spatial circle. More precisely, $w$ is an integer parametrizing the spectrally-flowed representations of $SL(2,R)$~\cite{Maldacena:2000hw}. 
In our conventions $w < 0$ and the derivation of the spectrum is valid for both the discrete (short strings) and continuous (long strings) representations of $SL(2,R)$~\cite{Apolo:2018qpq}. However, it is not yet clear what the role of the discrete representation on the massless BTZ background is in the proposed holographic dual.\footnote{We thank Soumangsu Chakraborty and David Kutasov for discussions on this point.}

We now show that the string theory spectrum~\eqref{ogspectrumleft} --~\eqref{ogspectrumu1} matches the spectrum of $J\bar T$-deformed CFTs~\eqref{sugawara} --~\eqref{qft_spectrumQ}. First, by subtracting eq.~\eqref{ogspectrumleft} from \eqref{ogspectrumright} we find that the angular momentum $E_L - E_R$ is independent of the deformation parameter $\l$. The charge flow eq.~\eqref{ogspectrumu1} also agrees with its $J\bar{T}$ counterpart~\eqref{qft_spectrumQ} after using eq.~\eqref{mulambda}. Finally, choosing 
  \eq{
  w = -1,\qquad \ell =R, \label{dictionary}
  }
  it is not difficult to show that eq.~\eqref{ogspectrumright} can be written as eq.~\eqref{sugawara}, which is the statement that the Sugawara combination of left-moving charges is invariant under the deformation. Note that while this is an important assumption in the field theory derivation of the spectrum in~\cite{Chakraborty:2018vja}, the $\l$-independence of the Sugawara combination of charges is a natural outcome of the string theory analysis.

%%%%%%%%%%%%%%%%%%%%%%%%%%%%%%%%%%%%%%%%%%%%%%

%%%%%%%%%%%%%%%%%%%%%%%%%%%%%%%%%%%%%%%%%%%%%%
\subsection{The conformal weight from the worldsheet} \label{se:conformalweight}
%%%%%%%%%%%%%%%%%%%%%%%%%%%%%%%%%%%%%%%%%%%%%%

The spectrum of strings on the warped background was derived on the spatial circle
  \eq{
  (u,v) \sim (u + 2 \pi, v + 2\pi),
  }
meaning that the dual $J\bar{T}$-deformed CFT is defined on a cylinder of size $2\pi \ell$. It is natural to ask what the fate of the deformed spectrum is when the dual CFT is defined on the plane. Since the $J\bar{T}$ deformation is expected to preserve a left-moving copy of the Virasoro algebra, we can equivalently ask what the corresponding conformal weight is in the deformed theory. 

String theory on AdS$_3$ features vertex operators that are identified with the left and right-moving components of the stress tensor of the dual CFT~\cite{Kutasov:1999xu}. A crucial ingredient in the construction of these vertex operators are the $SL(2,R)_L \times \widebar{SL(2,R)}_R$ currents of the worldsheet action. The latter can be shown to correspond to the $L_0$, $L_{\pm}$, and $\bar{L}_0$, $\bar{L}_{\pm}$ global symmetry generators of the dual CFT's left and right-moving Virasoro algebras~\cite{Giveon:1998ns,Kutasov:1999xu,deBoer:1998gyt}. More generally, an asymptotic symmetry generated by an asymptotic Killing vector $\xi$ accompanied by a gauge transformation parameter $\Lambda$ is given by the following vertex operator 
  \eq{
   -\frac{i}{\pi} \int dz^2  \big[ \L_{\xi} (g_{\mu\nu} + B_{\mu\nu}) + \p_{\mu} \ll_{\nu} - \p_{\nu} \ll_{\mu} \big] \p X^\mu \bar \p X^\nu.
  }
After the deformation, the $SL(2,R)_L \times \widebar{SL(2,R)}_R$ symmetries  are broken down to $SL(2,R)_L \times \widebar{U(1)}_R$ and we expect the left-moving Virasoro algebra of the dual CFT to be preserved by the deformation. Indeed, both the undeformed and deformed theories feature a local vertex operator that is identified with the left-moving component of the stress tensor in the dual CFT. This vertex operator is given by~\cite{Azeyanagi:2012zd}
   \eq{
   {T}({m})= - {k\over2\pi \ell}\oint \bigg\{ r \big[{\p} v + \l \big( \p \chi + \cos\tt \p\phi \big) \big]- \frac{i {m} {\p}r}{r} \bigg \} e^{i {m} u},\label{rightspacetimevirasoro}
  }
where $\oint \equiv \int_0^{2\pi} d\s$. It is not difficult to show that $T(m)$ satisfies the Virasoro algebra with central charge $c_{st} = 6 k I$ where $I$ is an operator that takes different values in different sectors of the theory, e.g.~short vs.~long string sectors~\cite{Giveon:2001up}. 
 
The left-moving conformal weight is defined by the zero mode of $T(m)$, namely,
  \eq{
  h - \frac{c_{st}}{24} \equiv  \Vev{L_0} = - \ell {T}(0) = \frac{k}{2\pi} \oint r \big[{\p} v + \l \big( \p \chi + \cos\tt \p\phi \big) \big] = \frac{1}{2\pi} \oint {j}_{\p_u}^\tau = \ell E_L, \label{conformalweightstring}
  }
where ${j}_{\p_u}$ is the current generating translations along $u$. Eq.~\eqref{conformalweightstring} is the usual relationship between the left-moving energy and conformal weight of a CFT defined on the cylinder. This expression is valid for any $\l$ and therefore applies to both the deformed and undeformed theories. Using eqs.~\eqref{ogspectrumleft} --~\eqref{ogspectrumu1} we find that the conformal weight satisfies
  \eq{
  {h}(\l) = {h}(0) - \frac{\ell \l }{w k} Q(0) E_R(\l) - \frac{\ell^2 \l^2 }{w k} E_R(\l)^2.
  }
This is the same deformed conformal weight derived in $J\bar T$-deformed CFTs, cf.~eq.~\eqref{defconformalweight}, where $w = -1$, $\ell \l = \mu k$, and $\ell = R$. In particular, in the $R \to \infty$ limit this expression reduces to eq.~\eqref{defconformalweight2} and agrees with the field theory derivation of ref.~\cite{Guica:2019vnb}.

%%%%%%%%%%%%%%%%%%%%%%%%%%%%%%%%%%%%%%%%%%%%%% 

%%%%%%%%%%%%%%%%%%%%%%%%%%%%%%%%%%%%%%%%%%%%%% 
\section{Warped BTZ and $J\bar{T}$ at finite temperature} \label{se:warpedbtz}
%%%%%%%%%%%%%%%%%%%%%%%%%%%%%%%%%%%%%%%%%%%%%%

In this section we consider the thermal background relevant for the string theory description of the $J\bar{T}$ deformation at finite temperature. A natural candidate is the warped BTZ black string constructed in~\cite{Azeyanagi:2012zd}. The latter was obtained via a TsT transformation of the BTZ black string mirroring the construction of the zero temperature background considered in the previous section. We will show that this background is equivalent to an instantaneous deformation of the worldsheet action by a marginal operator consisting of the antisymmetric product of two Noether currents.

%%%%%%%%%%%%%%%%%%%%%%%%%%%%%%%%%%%%%%%%%%%%%% 
\subsection{Warped BTZ from TsT transformations}
%%%%%%%%%%%%%%%%%%%%%%%%%%%%%%%%%%%%%%%%%%%%%% 

Let us begin by considering string theory on the BTZ black string. In our conventions the metric is given by
  \eq{
 {ds^2}{} &=  \ell^2 \bigg \{ {dr^2\over 4 \big (r^2 - 4 T_u^2 T_v^2 \big )} + r du dv + T_u^2 du^2 + T_v^2 dv^2 + d\Omega_3^2  \bigg \}, \label{btzmetric}
  }
where the $B$ field and dilaton are given in eqs.~\eqref{mbtzbfield} and~\eqref{mbtzdilaton}, and the $(u,v)$ coordinates satisfy
  \eq{
  (u,v) \sim (u + 2\pi, v + 2\pi). \label{btzidentification}
  }
The dimensionless temperatures $T_u$ and $T_v$ are related to the dimensionless left and right-moving energies of the BTZ background by
  \eq{
  {\cal E}_L= \frac{c_{st}}{6} {T_u^2}, \qquad {\cal E}_R =\frac{c_{st}}{6} {T_v^2}. \label{btzenergies}
  }
where $c_{st} = 6 kp$ is the central charge of the dual CFT with $k$ and $p$ denoting the magnetic and electric charges of the black string,
  \eq{
  \frac{1}{4\pi^2 \ell_s^2} \int_{S^3} H = k, \qquad \frac{\ell_s^2}{4\pi^2 G_6} \int_{S^3} e^{-2\Phi} \star H = p.
  }

The warped BTZ black string can be generated from the BTZ background via a TsT transformation, i.e.~by first T-dualizing along $\chi$, shifting $v' \to t^- + \frac{2 \l}{k} \psi$ and $\chi' \to \psi$ where $(v', \chi')$ denote the T-dual coordinates, and T-dualizing once more along $\psi$. Note that $t^-$ and $\psi$ denote the coordinates of the TsT-transformed background, and that this is the same TsT transformation responsible for the warped massless BTZ background reviewed in the previous section. In analogy with the zero temperature solution, we can perform an additional shift of coordinates 
  \eq{
  t^- = v + \frac{\l}{2} \chi, \qquad \psi = \chi + 2 \l T_v^2 v,  \label{shift22}
  }
to put the solution in a form where the local $SL(2,R)_L\times \overline{U(1)}_R$ isometries of the background are manifest. After the TsTs transformation, we obtain the following solution to the supergravity equations of motion, 
 \eq{
\!\!\!ds^2_{st} &=  \ell^2 \bigg \{ {dr^2\over 4 \big (r^2 - 4 T_u^2 T_v^2 \big )}\! + \big( r du + 2 T_v^2 dv \big) \Big( dv + \frac{\a}{2 T_v} \sigma_3 \Big)\! - T_v^2 dv^2\! + T_u^2 du^2\! + d\Omega_3^2  \bigg \} \label{btzmetricdef},  \\
 B & = {\ell^2 \over4} \bigg \{ \cos \t\, d\phi \we d \chi - 2 \big( r du + 2 T_v^2 dv \big) \we \Big ( dv +  \frac{\a}{2 T_v} \sigma_3 \Big ) + 2 \l T_v^2 dv \we d\chi \bigg\}, \label{btzbfielddef}
  }
where $ds_{st}$ denotes the line element in the string frame, $\sigma_3 =  d \chi + \cos\t\, d\phi$, and $\a$ satisfies
  \eq{
   \a = \frac{2\l T_v^2}{1 + \l^2 T_v^2}. \label{alphadef}
  }
Eqs.~\eqref{btzmetricdef} and~\eqref{btzbfielddef} are solutions of the equations of motion for any constant value of the dilation. For the TsT background, the dilaton is determined by requiring that $e^{-2\Phi} \sqrt{-g}$ is invariant under the transformation such that
  \eq{
  e^{-2\Phi} &= 1 + \l^2 T_v^2. \label{btzdilatondef} 
  }

While the coordinate transformation~\eqref{shift22} does not change the local geometry, it does affect its global properties and can lead to physically distinct solutions.  In other words, the background fields~\eqref{btzmetricdef} and~\eqref{btzbfielddef} describe different solutions depending on how we compactify the $u$, $v$, and $\chi$ coordinates. One possibility is to keep track of the TsT transformation and consider
  \eq{
		\textrm{internal circle:} &\quad  (u,v\!, \chi)  \sim \bigg (u, v - \frac{2\pi \l}{1 - \l^2 T_v^2}, \chi + \frac{4\pi}{1 - \l^2 T_v^2} \bigg), \label{tstidentificationv} \\
	\textrm{spatial circle:} &\quad  (u,v, \chi)  \sim \bigg (u + 2\pi, v + \frac{2 \pi}{1 - \l^2 T_v^2}, \chi - \frac{4\pi \l T_v^2}{1 - \l^2 T_v^2} \bigg). \label{tstcirclev} 
	}
The identifications~\eqref{tstidentificationv} and~\eqref{tstcirclev} are the natural ones to impose after the TsT transformation, which respectively correspond to 
  \eq{
  \psi \sim \psi + 4\pi, \qquad (u,t^-) \sim (u + 2\pi, t^- + 2\pi). 
  }
The thermodynamics and asymptotic symmetries of the TsT-transformed solution where $u+t^-$ is identified with the spatial direction have been previously studied in~\cite{Compere:2013bya,Song:2011sr,Azeyanagi:2012zd}. Notably, the entropy density per unit length along the $u+t^-$ direction, as well as the entropy with $u+t^-$ compactified~\eqref{tstcirclev}, remain unchanged after the TsT transformation. This provides a hint that we should compactify the coordinates via eqs.~\eqref{tstidentificationv} and~\eqref{tstcirclev} as the entropy of $J\bar T$-deformed CFTs is also unchanged by the deformation. Henceforth, we will adopt the TsT compactification of coordinates and use it to show that the deformed worldsheet spectrum, the entropy of the warped BTZ background, and the background charges match the corresponding expressions found in $J\bar{T}$-deformed CFTs.

%%%%%%%%%%%%%%%%%%%%%%%%%%%%%%%%%%%%%%%%%%%%%% 

%%%%%%%%%%%%%%%%%%%%%%%%%%%%%%%%%%%%%%%%%%%%%% 
\subsection{Worldsheet deformation} \label{suse:btzmarginal}
%%%%%%%%%%%%%%%%%%%%%%%%%%%%%%%%%%%%%%%%%%%%%% 

We now consider string theory on the warped BTZ background and give a worldsheet interpretation to the TsT transformation described in the previous subsection. In the TsT coordinates $t^-$ and $\psi$, the warped BTZ background reads
 \eq{
 \begin{split}
ds^2_{st} &=  \ell^2 \bigg \{ {dr^2\over 4 \big (r^2 - 4 T_u^2 T_v^2 \big )} + \frac{r du}{2(1+\l^2 T_v^2)}  \big[ 2 dt^- + \l (d\psi + 2 \cos\t\,d \phi ) \big]  + T_u^2 du^2  \\
& \hspace{1.2cm} +   d\Omega_3^2 + \frac{T_v^2 dt^- ( dt^- + \l \cos\t\, d\phi)}{1+\l^2 T_v^2}  - \frac{\l^2 T_v^2 ( d\psi + 2 \cos\t\,d\phi) d\psi}{4(1 + \l^2 T_v^2)}    \bigg \} ,
\end{split} \label{tstmetric}  \\
\begin{split}
 B & = \frac{\ell^2}{4(1+\l^2 T_v^2)} \bigg \{  \cos \t\, d\phi \we d \psi - r du \we \Big[ 2 dt^- +  \l (d \psi + 2 \cos\t\, d\phi ) \Big] \\
 & \hspace{3cm} - 2\l T_v^2 dt^- \we (d\psi + \cos\t\,d\phi) \bigg\},
 \end{split}\label{tstbfield}\\
 e^{-2\Phi} &= 1 + \l^2 T_v^2,
  }
which makes manifest one of the advantages of using the TsTs coordinates $v$ and $\chi$, as the metric~\eqref{btzmetricdef} and $B$-field~\eqref{btzbfielddef} simplify significantly in terms of these variables. In particular, the deformed worldsheet action is given by
\eq{
   \begin{split}
 S_\lambda = k \! \int \!dz^2 \bigg\{ &\frac{\p r \bar{\p} r}{4 \big(r^2 - 4 T_u^2 T_v^2 \big)} + \frac{r  \big[ 2 \p t^- + \l \big( \p \psi + 2 \cos\t\, \p\phi \big) \big]\bp u }{2(1 + \l^2 T_v^2)} +\frac{T_v^2   \p t^- \bp t^-}{1 + \l^2 T_v^2}  \\
 & + T_u^2 \p u \bp u + \frac{\cos\t\,\p\phi\big(\bp \psi +2 \l T_v^2 \bp t^- \big)}{2(1 + \l^2 T_v^2)}   - \frac{\l T_v^2\big( \p t^- \bp \psi - \p \psi \bp t^- \big)}{2(1 + \l^2 T_v^2)} \\
 &+ \frac{\p\psi \bp \psi}{4(1 + \l^2 T_v^2)}     + \frac{1}{4}   \p\t\bp\t +  \frac{1}{4}\p\phi\bp\phi \bigg\}.
  \end{split} \label{btzactiondef1}
  }
 In the $T_v \to 0$ limit, the deformed action~\eqref{btzactiondef1} reduces to its zero temperature counterpart described in Section~\ref{se:jtbarspectrum}. This suggests a  connection to the single-trace $J\bar{T}$ deformation of the dual CFT.
To sharpen this connection, let us introduce the Noether currents $j_{\p_{t^-}}$ and $j_{\xi_{U(1)}}$ generating translations along the Killing vectors $\p_{t^-}$ and $ \xi_{U(1)} =2\p_{\psi} - \l \p_{t^-}$. These currents are defined in eqs.~\eqref{jtm} and~\eqref{ju1}, respectively, and their zero mode charges correspond to (minus) the deformed right-moving energy $-E_R(\l)$ and the deformed $U(1)$ charge $Q(\l)$ (see Section~\ref{se:worldsheet}). In terms of these Noether currents, we find that the deformed action $S_\l$ satisfies the following differential equation
   \eq{
   \frac{\p S_{\l} }{\p \l} =  \frac{1}{k} \int d^2z\,  (j_{\p_{t^{-}}})^a  (j_{{\xi_{U(1)}}})^b \e_{ab}, \label{diffdef1}
   }
where $\e_{z\bar{z}} = - \e_{\bar{z}z} = 1$ is the Levi-Civita symbol. Thus, the TsT transformation is equivalent to an instantaneous deformation of the worldsheet action by the antisymmetric product of two Noether currents. Eq.~\eqref{diffdef1} mirrors the definition of the $J\bar{T}$ deformation~\eqref{jtbardeformation} in the dual field theory, the latter of which is also instantaneous and the antisymmetric product of $J_{\mu}$ and $T_{\mu \bar{x}}$, whose zero modes correspond to the deformed $U(1)$ charge and the right-moving energy, respectively.

 Concrete evidence for the correspondence between the marginal deformation~\eqref{diffdef1} and the single-trace $J\bar{T}$-deformation of the dual CFT is given by the worldsheet spectrum derived in Section~\ref{se:worldsheet}, as well as the gravitational charges and entropy of the warped BTZ background computed in Section~\ref{se:entropy} --- quantities that match the corresponding expressions in $J\bar{T}$-deformed CFTs.
 
 It is interesting to note that the warped BTZ black string considered in this paper is closely related to another background whose worldsheet interpretation is that of an exactly marginal deformation by the product of two chiral currents. As shown in detail in Appendix~\ref{se:currentcurrent}, this background is locally equivalent to the the warped BTZ black string considered in this paper and is obtained by an additional change of coordinates and a gauge transformation. Crucially, the gauge transformation is not compatible with the TsT identification of coordinates and as a result this background differs from the warped BTZ black string on its global properties. The latter are essential for the connection to $J\bar{T}$ deformations and it is only on the TsT background where we can reproduce the spectrum and thermodynamics of $J\bar{T}$-deformed CFTs.

%\bigskip

To summarize, we propose that the holographic dual to $J\bar{T}$-deformed CFTs at finite temperature can be described by the NS-NS sector of type IIB string theory on the warped BTZ background described, in different coordinate systems, in eqs.~\eqref{btzmetricdef}, \eqref{btzbfielddef} and~\eqref{tstmetric}, \eqref{tstbfield}. This background is obtained from a TsT transformation of the undeformed BTZ black string and we have shown that it can be interpreted as an instantaneous deformation of the worldsheet by a marginal operator.

%%%%%%%%%%%%%%%%%%%%%%%%%%%%%%%%%%%%%%%%%%%%%% 

%%%%%%%%%%%%%%%%%%%%%%%%%%%%%%%%%%%%%%%%%%%%%%
\section{Strings on warped BTZ}  \label{se:worldsheet}
%%%%%%%%%%%%%%%%%%%%%%%%%%%%%%%%%%%%%%%%%%%%%%
 
In this section we derive general expressions for the spectrum of string theory on the warped BTZ background that are valid for generic compactifications of the $u$, $v$, and $\chi$ coordinates. We then consider the TsT compactification and show that the spectrum of strings winding along the spatial circle matches the spectrum of $J\bar{T}$-deformed CFTs. 

%%%%%%%%%%%%%%%%%%%%%%%%%%%%%%%%%%%%%%%%%%%%%%
\subsection{Noether currents and winding strings}   \label{se:worldsheetcurrents}
%%%%%%%%%%%%%%%%%%%%%%%%%%%%%%%%%%%%%%%%%%%%%%

We begin by describing the Noether currents and conserved charges of the deformed worldsheet action. For convenience we work with the $v$ and $\chi$ coordinates introduced in Section~\ref{se:warpedbtz} which simplify the analysis and make the preserved symmetries of the theory more transparent.\footnote{For completeness, in terms of the $v$ and $\chi$ coordinates, the worldsheet action is given by
   \eqsp{
  S_{\l} = k \int dz^2 \bigg\{ & \frac{\p r \bar{\p} r}{4 \big(r^2 - 4 T_u^2 T_v^2 \big)} + \big( r \bp u + 2T_v^2 \bp v\big) \bigg[ \p v + \frac{\a}{2 T_v} \big ( \p\chi + \cos\t\,\p\phi \big) \bigg ] - T_v^2 \p v\bar{\p} v + T_u^2 \p u \bar{\p} u \\
  &+ \frac{\l T_v^2}{2}  (\p v \bar{\p} \psi - \bar{\p} v\p \psi) + \frac{1}{4} \Big [\big(\p\chi + 2 \cos\t\,\p\phi \big) \bp \chi + \p\t\bp\t + \p\phi\bp\phi \Big] \bigg \}. \notag
  }
  \vspace{-30pt}
  } 
Before the deformation, the theory features locally conserved currents satisfying left and right-moving $SL(2,R) \times SU(2)$ Kac-Moody algebras. However, in contrast to the global and Poincar\'e AdS$_3$ backgrounds, not all of the $SL(2,R)$ currents are globally well-defined with the exception of
  \eq{
  j^1(z) = k \bigg( \frac{r \p v}{2T_u} + T_u \p u \bigg), \qquad \bar{j}^1(\bar{z}) =  k \bigg(  \frac{r \bp u}{2T_v} + T_v \bp v \bigg), \label{j10}
  }   
where the superscript means that these currents are spacelike. We note that $j^1$ and $\bar{j}^1$ are the same chiral currents obtained from the WZW formulation of the worldsheet action. These currents are closely related to the Noether currents generating translations along $u$ and $v$ but differ from the latter by divergence-free terms. When the temperatures vanish, the $j^1$ and $\bar{j}^1$ currents become null and reduce to the expressions featured in Section~\ref{se:jtbar},
  \eq{
  \lim_{T_u \to 0} (2T_u) j^1 = j^-, \qquad  \lim_{T_v \to 0} (2T_v) \bar{j}^1 = \bar{j}^-. \label{limitcurrents}
  }

After the deformation the currents associated to translations along $u$, $v$, and $\chi$ satisfy
  \eq{
    j_{\p_u} &= \bigg( 2T_u j^1 + \frac{\a}{T_v} r k^3 \bigg) \bp + k T_u^2 \t_{(u)}, \label{j1k30} \\
  j_{\p_v} &=   2T_v \big( \bar{j}^1 \p +  \a k^3 \bp \big) - k T_v^2 \t_{(v-\l \chi/2)}, \label{j1k31}\\
  j_{\p_\chi} &=  \a \bar{j}^1  \p + {k}^3 \bp + \frac{k}{4} \t_{(\chi - 2 \l T_v^2 v)}, \label{j1k32}
  }
  where $j_\xi$ denotes the Noether current corresponding to the background Killing vector $\xi$, $\t_{(f)}$ is a divergence free current defined by
 \eq{
\t_{(f)} \equiv \bp f\, \p - \p f\, \bp,
 }
and $(j^1, \bar{j}^1, {k}^3)$ are the $SL(2,R)$ and $SU(2)$ currents given in eqs.~\eqref{k30} and~\eqref{j10} which remain chirally conserved in the deformed theory.

The $\t_{(f)}$ terms featured in eqs.~\eqref{j1k30} --~\eqref{j1k32} contribute to the Noether charges in topologically nontrivial sectors of the theory, e.g.~in the winding string sector. This is a property of string theory that is already present for strings winding on the undeformed BTZ background~\cite{Hemming:2001we,Rangamani:2007fz}. In this paper, we will consider strings winding along the $u$, $v$, and $\chi$ coordinates,
  \eq{
  \oint \p_{\s} u = {2\pi w_u} , \qquad \oint \p_{\s} v =  2\pi w_v , \qquad   \oint \p_{\s} \chi = 4\pi  w_{\chi},\label{windingbc2}
  }
whereupon the zero modes of the $\t_{(f)}$ currents become proportional to the winding. The values of the winding parameters $w_u$, $w_v$, and $w_{\chi}$ depend on the global properties of the warped BTZ spacetime. The spectrum derived in the next subsection is valid for arbitrary values of the winding and can be applied to backgrounds satisfying different compactifications of coordinates. We will focus on the TsT compactification in Section~\ref{suse:matching}.

The zero modes of the $j_{\p_u}$, $j_{\p_v}$, and $j_{\p_{\chi}}$ currents yield the physical momenta of the string along the $u$, $v$, and $\chi$ coordinates. However, we find that the derivation of the spectrum simplifies significantly in terms of the ``winding momenta". The latter are defined as the zero mode charges of the Noether currents $j_{\chi}$ modulo the $\t_{(f)}$ terms --- conserved currents we denote by $j'_{\xi}$ --- such that
  \eq{
  {p}_w &\equiv \frac{1}{2\pi} \oint j'^{\,\tau}_{\p_{u}}, \qquad  -{\bar{p}_w} \equiv \frac{1}{2\pi} \oint {j}'^{\,\tau}_{\p_v}, \qquad \frac{{q}_w}{2} \equiv \frac{1}{2\pi} \oint {j}'^{\,\tau}_{\p_{\chi}}, \label{zeromodecharges}
  }
where $j_{\xi}^{\tau} = j^z_{\xi} + j^{\bar{z}}_{\xi}$ and the normalization of $q_w$ follows from the fact that $\chi \sim \chi + 4 \pi$ in the undeformed theory.

%%%%%%%%%%%%%%%%%%%%%%%%%%%%%%%%%%%%%%%%%%%%%%

%%%%%%%%%%%%%%%%%%%%%%%%%%%%%%%%%%%%%%%%%%%%%%
\subsection{Spectrum via spectral flow} \label{suse:wbtzspectrum}
%%%%%%%%%%%%%%%%%%%%%%%%%%%%%%%%%%%%%%%%%%%%%%

A crucial feature of the TsT transformation is that the deformed equations of motion and constraints can be obtained from the corresponding expressions in the \emph{undeformed} theory after a nonlocal change of coordinates~\cite{Azeyanagi:2012zd}. Let us continue to denote the coordinates of the deformed theory by $x^{\mu}$ and use $\hat{x}^{\mu}$ to denote the coordinates of the undeformed one. The aforementioned change of coordinates affects only $v$ and $\chi$ and satisfies
  \eq{
  \p\hat{v} &= \p v + \frac{\a}{ k T_v} k^3, \qquad \quad \,\,\bp\hat{v} = \bp v - \frac{\l \a}{k} \bar{j}^1,  \label{vhat1}\\
  \p\hat{\chi} &= \p\chi - \frac{2 \l \a T_v}{k} k^3, \qquad \bp\hat{\chi} = \bp\chi + \frac{2\a}{k} \bar{j}^1.   \label{chihat2}
  }
Eqs.~\eqref{vhat1} and~\eqref{chihat2} allow us to relate the chirally conserved currents before and after the deformation. In particular, we are interested in the chiral currents corresponding to the Cartan generators of $\widebar{SL(2,R)}_R \times SU(2)_L$ in the undeformed theory, the latter of which can be written in terms of deformed variables as
  \eq{
  \hat{\bar{j}}^1  & \equiv k \bigg ( \frac{\hat{r} \bp \hat{u}}{2T_v} +  T_v \bp \hat{v} \bigg )  =  \sqrt{1 - \a^2} \bar{j}^1, \qquad  \hat{{k}}^3  \equiv \frac{k}{2} \big ( \p \hat{\chi} + \cos\hat{\t} \p \hat{\phi} \big )  = \sqrt{1 - \a^2} {k}^3. \label{hatk3}
  }
We furthermore note that, up to topological terms, the $\hat{k}^3$ current  can be expressed as the Noether current of a Killing vector ${\xi_{U(1)}}$ defined by
  \eq{
     {j}_{\xi_{U(1)}} &=2 \hat{k}^3+ \frac{k}{2} \t_{(\chi + 2\l T_v^2 v)}, \qquad {\xi_{U(1)}} = \frac{2}{\sqrt{1 - \a^2}} \Big ( \p_{\chi} - \frac{\a}{2T_v} \p_{v} \Big ) \label{ju1}
  }
This will be useful when we compute the gravitational charges of the warped BTZ background in Section~\ref{se:entropy}.

We will identify the deformed $U(1)$ charge with the zero mode of the chiral $\hat{k}^3$ current given in eq.~\eqref{hatk3} (up to a factor of 2). An important assumption in the field theory analysis of the $J\bar{T}$ deformation is that the deformed $U(1)$ current remains chiral along the flow~\cite{Chakraborty:2018vja}. We will assume that this property extends to the corresponding current in the worldsheet theory, an assumption that played an important role in the string theory derivation of the spectrum in~\cite{Apolo:2018qpq}. In particular, the $U(1)$ charge of the deformed theory can be written in terms of the winding momenta $\bar{p}_w$ and $q_w$ as
  \eq{
{{Q}(\lambda)} \equiv \frac{1}{\pi} \oint \hat{k}^3 =  \frac{1}{\sqrt{1-\a^2}} \bigg(q_w + \frac{\a \bar{p}_w}{T_v} \bigg) \label{U1charge}.
  }
  Interestingly, the $T$ component of the stress tensor is also assumed to remain chiral along the flow in the field theory analysis of the $J\bar{T}$ deformation~\cite{Chakraborty:2018vja}. Up to a topological term, this is also a property of the worldsheet current $j_{\p_u}$ generating translations along $u$~\eqref{j1k30}. 
  
A consequence of the nonlocal change of coordinates is that the $\hat{x}^{\mu}$ variables obey nontrivial boundary conditions. Indeed, integrating $\oint \p_{\s} \hat{x}^{\mu}$ yields
  \eq{
\hat{x}^{\mu}(\s + 2 \pi) = \hat{x}^{\mu}(\s) + 2 \pi \g_{x^{\mu}}, \label{twistedbc}
  }
where the only nonzero $\g_{x^{\mu}}$ parameters are given by
 \eq{
   \g_{u} & = w_u , \label{gammau} \\
 \g_v &= w_v + \frac{ \l \a }{k} \frac{1}{2\pi} \oint \bar{j}^1 + \frac{\a}{k T_v} \frac{1}{2\pi} \oint {k}^3= w_v + \frac{\lambda\big ({q}_w  + \l \bar{p}_w \big )}{k \big({1 - \lambda^2T_v^2} \big)} ,\label{gammav}\\
   \g_{\chi} &= 2  w_{\chi}  - \frac{2\a}{k} \frac{1}{2\pi} \oint \bar{j}^1 - \frac{2 \l {\a}{T_v}}{k} \frac{1}{2\pi} \oint {k}^3= 2  w_{\chi} + \frac{2\lambda\big ( \bar{p}_w + \l T_v^2 {q}_w \big )}{k \big(1-\lambda^2T_v^2 \big)}  . \label{gammachi}
  } 
The worldsheet integrals featured in eqs.~\eqref{gammav} and~\eqref{gammachi} imply that the $\hat{v}$ and $\hat{\chi}$ coordinates satisfy nonlocal boundary conditions that depend on the $\bar{p}_w$ and $q_w$ momenta of the string. Thus, any solution to string theory in the warped BTZ background is equivalent to a solution of the undeformed theory satisfying momentum-dependent boundary conditions. Indeed, we note that the classical string solutions given in Appendix~\ref{se:classicalstrings} can also be obtained from the nonlocal change of coordinates in eqs.~\eqref{vhat1} and~\eqref{chihat2}. In this case, $\hat{x}^{\mu}$ describes a geodesic in the undeformed BTZ background satisfying the boundary conditions in eq.~\eqref{twistedbc}. Solving eqs.~\eqref{vhat1} and~\eqref{chihat2} then yields the winding strings described in Appendix~\ref{se:classicalstrings}. At the quantum level, the boundary conditions are enforced by dressing vertex operators with nonzero $\bar{p}_w$ and $q_w$ momenta by the appropriate spectral flow operators~\cite{Azeyanagi:2012zd}.

We now have the necessary ingredients to derive the spectrum of strings on the warped BTZ background. The twisted boundary conditions~\eqref{windingbc2} induce a spectral flow transformation on the undeformed theory. In order to see this we note that the twisted boundary conditions can be obtained from the following shift of coordinates
  \eq{
 \hat{u} = \til{u} + \g_u {z}, \qquad \hat{v} = \til{v} - \g_v \bar{z}, \qquad \hat{\chi} = \til{\chi} + \g_{\chi} {z}, \label{shift}
  }
where $\til{x}^{\mu}$ are coordinates in the undeformed theory satisfying trivial boundary conditions, i.e.~$\oint \p_{\s} \til{x}^{\mu} = 0$. The change of coordinates~\eqref{shift} leads to shifts and rescalings of the conserved currents that are equivalent to spectral flow. In particular, the currents associated to the Cartan generators of $SL(2,R)_L \times \widebar{SL(2,R)}_R \times SU(2)_L$ shift according to
  \eq{
  \til{{j}}^1 &\to \hat{{j}}^1 = \til{{j}}^1 + k T_u \g_u, \qquad \til{\bar{j}}^{1} \to \hat{\bar{j}}^{1} = \til{\bar{j}}^{1} - k T_v \g_v, \qquad \til{{k}}^3 \to \hat{{k}}^3 = \til{{k}}^3 + \frac{k}{2} \g_{\chi},
  }
where the tilded currents are given by eqs.~\eqref{k30} and~\eqref{j10} with ${x}^{\mu}$ replaced by $\til{x}^{\mu}$. \linebreak[2] In terms of the winding  momenta, the zero modes of the $ \til{{j}}^1$ and $ \til{\bar{j}}^1$ currents are given by
   \eq{
    \frac{1}{2\pi } \oint \til{{j}}^1 & = \frac{{p_w}}{2 T_u} - k w_u T_u, \qquad  \frac{1}{2\pi } \oint \til{\bar{j}}^1 = -\frac{\bar{p}_w}{2 T_v} + k w_v T_v,
    }
while the zero mode of $\til{k}^3$ is, by definition, the $U(1)$ charge of the undeformed theory
  \eq{
  \frac{q}{2} \equiv \frac{1}{2\pi} \oint \til{{k}}^3  &= \frac{{q_w}}{2} - k w_{\chi}. \label{undeformedU1}
  }
  %

%%%%%%%%%%%%%%%%%%%%%%%%%%%%%%%%%%%%%%%%%%%%%%
%%%%%%%%%%%%%%%%%%%%%%%%%%%%%%%%%%%%%%%%%%%%%%
%%%%%%%%%%%%%%%%%%%%%%%%%%%%%%%%%%%%%%%%%%%%%%
\begin{comment}
In the deformed theory, the left-moving component of the worldsheet stress tensor reads
  %
  \eq{
  \hat{T} &= -\frac{1}{k} \Big [ \big (\hat{j}^1 \big)^2 - \hat{j}^{+} \hat{j}^- + \big ( \hat{{k}}^3 \big )^2 - \hat{{k}}^+ \hat{{k}}^-  + \dots\Big ], \label{tmunu1}
  }
  %   
where we have assumed $k \gg 1$, $\hat{j}^{\pm}$ and $\hat{k}^{\pm}$ are the other currents of the $SL(2,R)_L \times SU(2)_L $ half of the symmetry group, and we have omitted the contributions from the ${\cal M}^4$. The right-moving component $\hat{\bar{T}}$ is obtained from~\eqref{tmunu1} by trading the unbarred currents for barred ones. Using the nonlocal change of coordinates it is not difficult to show that these expressions reproduce the stress-energy tensor of the deformed worldsheet action given in eq.~\eqref{leftstresstensor}. 
\end{comment}
%%%%%%%%%%%%%%%%%%%%%%%%%%%%%%%%%%%%%%%%%%%%%%
%%%%%%%%%%%%%%%%%%%%%%%%%%%%%%%%%%%%%%%%%%%%%%
%%%%%%%%%%%%%%%%%%%%%%%%%%%%%%%%%%%%%%%%%%%%%%

Although the symmetry algebra does not change under spectral flow, the chiral components of the stress-energy tensor shift according to
  \eq{
  \hat{T} &= \til{T} - 2 T_u \g_u \til{j}^1 - k \big(T_u \g_u\big)^2 - \g_{\chi} \til{k}^3- \frac{k}{4} \g_{\chi}^2, \\
  \hat{\bar{T}} &= \til{\bar{T}} + 2 T_v \g_v \til{\bar{j}}^1 - k \big( T_v \g_v \big)^2,
  }
where we have assumed $k \gg 1$ and $\til{T}$, $\til{\bar{T}}$ are the components of the stress tensor in the undeformed theory without winding boundary conditions. Thus, the zero modes of the deformed stress tensor $\hat{L}_{0} = -\frac{1}{2\pi} \oint \hat{T}$ and $\hat{\bar{L}}_{0} = -\frac{1}{2\pi} \oint \hat{\bar{T}}$ are given by
  \eq{
  \hat{L}_0 = &\til{L}_0 +  p_w  w_u - k ( T_u  w_u )^2  + q_w w_{\chi}  - k w_{\chi}^2  +   \frac{\a^2}{1 - \a^2} \frac{1}{k} \bigg ( \frac{\bar{p}_w^2}{4 T_v^2} + \frac{{q}_w^2}{4}  + \frac{ \bar{p}_w {q}_w}{2 \a T_v} \bigg),   \label{L0} \\
  \hat{\bar{L}}_0  = & \til{\bar{L}}_0 + \bar{p}_w w_v - k ( T_v  w_v)^2 +    \frac{\a^2}{1 - \a^2} \frac{1}{k} \bigg ( \frac{\bar{p}_w^2}{4 T_v^2} + \frac{{q}_w^2}{4}  + \frac{ \bar{p}_w {q}_w}{2 \a T_v} \bigg). \label{L0bar}
  }
where $\til{L}_0 = -\frac{1}{2\pi} \oint \til{{T}}$ and $\til{\bar{L}}_0 =  -\frac{1}{2\pi} \oint \til{\bar{T}}$. The latter satisfy
  \eq{
  \til{L}_0 \Ket{\psi} =  \Big( -\frac{j(j-1)}{k} + N + \dd \Big ) \Ket{\psi}, \qquad \til{\bar{L}}_0 \Ket{\psi}=  \Big (-\frac{\bar{j}(\bar{j}-1)}{k} + \bar{N} + \bar{\dd} \Big )\Ket{\psi}, \label{L0til}
  }
where $\Ket{\psi}$ is a state with $SL(2,R)$ weight $j$ at level $(N, \bar{N})$, and $(\dd, \bar{\dd})$ denote the contributions of the $S^3 \times {\cal M}^4$ parts of the background~\cite{Maldacena:2000hw}. 

Eqs.~\eqref{L0} and~\eqref{L0bar} extend the validity of the classical analysis of the spectrum carried out in Appendix~\ref{se:classicalstrings}. In particular, these equations reduce to the spectrum given in~\cite{Azeyanagi:2012zd} for strings with zero winding as well as the expressions given in~\cite{Apolo:2018qpq} for strings on the deformed zero temperature background provided that $w_u = w_v = w$ and $w_{\chi} = 0$. Imposing the Virasoro constraints on a state $\Ket{\psi}$,
  \eq{
  \big(\hat{L}_0 -1 \big) \Ket{\psi} = 0, \qquad \big(\hat{\bar{L}}_0-1 \big) \Ket{\psi} = 0, 
  }
yields the spectrum of the deformed theory. We see that for states with nonvanishing momenta $p_w$, $\bar{p}_w$, and $q_w$, the $\a$-dependent contributions to the spectrum are universal, as are the local contributions that originate from winding.

%%%%%%%%%%%%%%%%%%%%%%%%%%%%%%%%%%%%%%%%%%%%%%

%%%%%%%%%%%%%%%%%%%%%%%%%%%%%%%%%%%%%%%%%%%%%%
\subsection{Matching the spectrum to $J\bar{T}$} \label{suse:matching}
%%%%%%%%%%%%%%%%%%%%%%%%%%%%%%%%%%%%%%%%%%%%%%

We now show that the spectrum of strings winding on the warped BTZ background matches the spectrum of $J\bar T$-deformed CFTs. A few subtleties arise in the matching of the spectrum on the finite temperature background and it is not a priori clear what the appropriate map between variables in the worldsheet and the dual field theory is. In order to match the spectrum we need to understand (\emph{i})~which circle in the bulk corresponds to the spatial circle in the dual field theory. This subtlety arises only at finite temperature and hence does not affect the discussion of the warped massless BTZ background considered in Section~\ref{se:jtbar}; (\emph{ii})~what the deformed and undeformed $U(1)$ charges are; and (\emph{iii})~which Noether charges correspond to the left and right-moving energies in the dual field theory. We note that subtleties in the identification of the energy and angular momentum already arise for strings on the undeformed BTZ background~\cite{Hemming:2001we,Rangamani:2007fz}. In this paper we address these points as follows:
\vspace{-6pt}
\begin{itemize}
\item[(\emph{i})] we focus on the warped BTZ background with the TsT compactification of coordinates where the spatial circle is defined in eq.~\eqref{tstcirclev};

\vspace{-6pt}
\item[(\emph{ii})] we identify the undeformed $U(1)$ charge with the $SU(2)$ quantum number before spectral flow~\eqref{undeformedU1}, i.e.~$\frac{1}{\pi} \oint \tilde{k}^3$, whereas the deformed $U(1)$ charge is identified with the zero mode of the same current after spectral flow~\eqref{U1charge}, namely $\frac{1}{\pi} \oint \hat{k}^3$. This is the same identification of $U(1)$ charges used in the zero temperature case~\cite{Apolo:2018qpq}.

\vspace{-6pt}
\item[(\emph{iii})] we identify the left and right-moving energies with the zero modes of the Noether currents $j_{\p_u}$ and $-j_{\p_{t^-}}$. This is consistent with the analysis of string theory on the undeformed BTZ background carried out in~\cite{Hemming:2001we}. 
\end{itemize}
\vspace{-6pt}

The spectrum of strings winding on the warped BTZ background is sensitive to the global properties of the spacetime through the winding parameters  $w_u$, $w_v$, and $w_{\chi}$. For a string winding along the spatial circle~\eqref{tstcirclev} these parameters satisfy
  \eq{
  w_u = w, \qquad w_v = \frac{w}{ 1-\l^2 T_v^2} , \qquad w_{\chi} = -\frac{\l T_v^2 w}{ 1-\l^2 T_v^2}, \qquad w \in \mathbb{Z}. \label{tstwinding2}
  }
This choice guarantees that the string winds $w$ times along the spatial circle, namely
  \eq{
 \oint \p_{\s} u = 2 \pi w, \qquad \oint \p_{\s} t^- = 2\pi w, \qquad \oint \p_{\s} \psi= 0. \label{tstwindingbc}
  }
In particular, the Noether current corresponding to translations along $t^-$ is given by
  \eq{
  j_{\p_{t^-}} &= \frac{2T_v}{1 + \l^2 T_v^2} \big( \bar{j}^1 \bp + \l T_v {k}^3 {\p} \big) - k T_v^2 \t_{(v)}, \qquad \p_{t^-} = \frac{1}{1 - \l^2 T_v^2} \big ( \p_v - 2 \l T_v^2 \p_{\chi} \big), \label{jtm}
  }
whose zero mode charge corresponds to the momenta of the string along the $t^-$ coordinate.

In the derivation of the Noether currents we have been careful to keep track of the topological $\t_{(f)}$ terms, the latter of which can contribute to the Noether charges in the winding string sector. In the undeformed theory, the Noether currents also feature topological terms which distinguish them from the chiral currents obtained in the WZW formulation. As discussed in~\cite{Hemming:2001we,Rangamani:2007fz}, these topological terms are physically relevant. Therefore, we identify the left and right-moving energies of the $J\bar{T}$-deformed CFT with the zero mode charges of the $j_{\p_u}$ and $j_{\p_{t^-}}$ currents, which are the natural quantities to consider in the TsT-transformed background. We thus have
  \eq{
  \ell E_L(\l) &= \frac{1}{2\pi} \oint j_{\p_u}^\tau = p_w - k T_u^2 w, \label{EL} \\
  -\ell E_R(\l) & = \frac{1}{2\pi} \oint j_{\p_{t^-}}^\tau \!\!= -\frac{1}{1 - \l^2 T_v^2} \Big[ \bar{p}_w + \l T_v^2 q_w - k T_v^2 w \Big].
  }

In terms of the left and right-moving energies $E_L$ and $E_R$, the zero modes of the stress tensor in eqs.~\eqref{L0} and~\eqref{L0bar} are given by
  \eq{
  \hat{L}_0 = \til{L}_0 + w \ell E_L(\l) + \frac{1}{k} \l  \ell E_R(\l) \big[  \l \ell E_R(\l) + q \big],  \label{L0tst}\\
  \hat{\bar{L}}_0 = \til{\bar{L}}_0 + w \ell E_R(\l)  + \frac{1}{k}  \l \ell E_R(\l) \big [\l \ell E_R(\l) + q \big]. \label{L0bartst}
  }
By holding the eigenvalues of $\til{L}_0$ and $\til{\bar{L}}_0$ fixed before and after the deformation, we find that the left and right-moving energies of strings on the warped BTZ background satisfy the same formulae obtained in the zero temperature background, namely eqs.~\eqref{ogspectrumleft} and~\eqref{ogspectrumright}. It is not difficult to show that the undeformed and deformed $U(1)$ charges $q = Q(0)$ and $Q(\l)$ in eqs.~\eqref{undeformedU1} and~\eqref{U1charge} are related in the same way as in the zero temperature background~\eqref{ogspectrumu1}. 
 
Consequently, we find that the angular momentum and the Sugawara combination of left-moving charges $E_L + Q^2/4 \ell w k$ are left invariant by the deformation. Furthermore, matching the worldsheet and field theory parameters via eqs.~\eqref{mulambda} and~\eqref{dictionary}, we recover the spectrum of $J\bar{T}$-deformed CFTs given in eqs.~\eqref{sugawara} --~\eqref{qft_spectrumQ}. We emphasize that matching the $J\bar{T}$-deformed spectrum is only possible for the warped BTZ background satisfying the TsT compactification of coordinates. For other compactifications, the worldsheet spectrum features terms that are not found in the field theory derivation of the spectrum. 

Finally, we note that for the winding parameters given in eq.~\eqref{tstwinding2}, both the undeformed and deformed $U(1)$ charges can be related to Killing vectors on the warped BTZ background. For the deformed $U(1)$ charge this follows from the fact that the Noether current is chiral up to a topological term whose zero mode vanishes exactly when eq.~\eqref{tstwinding2} is satisfied, so that
  \eq{
  {Q(\l)} & \equiv \frac{1}{ \pi} \oint \hat{k}^3 = \frac{1}{2 \pi} \oint j^{\tau}_{\xi_{U(1)}}.\label{deformedU1}
  }
On the other hand, for the undeformed $U(1)$ charge $q$, explicit calculation shows that it is associated with translations along $\chi$, that is
  \eq{
  {q} &   = {q_w} + \frac{2k \l T_v^2 w}{1 - \l^2 T_v^2} = \frac{1}{ \pi} \oint j^{\tau}_{\p_{\chi}},
  }
The relationship between the Killing vectors of the warped BTZ background, the corresponding worldsheet currents, and the conserved charges in the dual field theory are summarized in Table~\ref{table}. 

Interestingly, since the $\p_{\chi}$ Killing vector differs from $\xi_{U(1)}$ by a term proportional to $\p_{t^-}$ we find that the differential equation~\eqref{diffdef1} can also be expressed as
   \eq{
   \frac{\p S_{\l}}{\p \l} = \frac{1}{k} \int d^2z\, (j_{\p_{t^{-}}})^a (j_{\xi_{U(1)}})^b \e_{ab}= \frac{2}{k} \int d^2z\, (j_{\p_{t^{-}}})^a (j_{\p_{\chi}})^b \e_{ab} .
   }
This indicates that the $U(1)$ current in the definition of the $J\bar T$ deformation in the dual field theory can be  associated with either $\xi_{U(1)}$ or $2\p_\chi$. The former remains chiral  along the flow but carries a deformed charge $Q(\lambda)$, while the latter is nonchiral but has a fixed charge $q$. This is consistent with the observations  made in ref.~\cite{LeFloch:2019rut} from a purely field theoretical approach.

	\renewcommand{\arraystretch}{1.25}
  \begin{table}[h!]
\begin{center}
\begin{tabular}{c|c|c}
Killing vector & worldsheet current & $J\bar T$-deformed CFT \\ \hline
$ \p_u$&$ j_u$&$E_L(\l) $\\
$-\p_{t^-}$&$ -j_{t^-}$ &$E_R(\l) $\\
2$\p_\chi$ & $2 \tilde{k}^3$ & $Q(0) = q$\\
$\xi_{U(1)}$&$2 \hat{k}^3$&$Q(\lambda)$\\
\end{tabular}
  \caption{Killing vectors and their corresponding worldsheet currents and field theory charges}
  \label{table}
\end{center}
  \end{table}

%\bigskip 

To recap, with the choice of spatial circle~\eqref{tstcirclev} and the identification of charges~\eqref{EL} --~\eqref{deformedU1} we find that the spectrum of strings winding on the warped BTZ background matches the spectrum of $J\bar{T}$-deformed CFTs, with a dictionary summarized in Table~\ref{table}. In particular, we noted that topological terms in the Noether currents (which are also present before the deformation) play an important role in the derivation of the spectrum. This result provides additional evidence for our identification of the warped BTZ black string as a thermal background dual to $J\bar{T}$-deformed CFTs at finite temperature.

%%%%%%%%%%%%%%%%%%%%%%%%%%%%%%%%%%%%%%%%%%%%%%

%%%%%%%%%%%%%%%%%%%%%%%%%%%%%%%%%%%%%%%%%%%%%%
\section{Thermodynamics of Warped BTZ} \label{se:entropy}
%%%%%%%%%%%%%%%%%%%%%%%%%%%%%%%%%%%%%%%%%%%%%%

We now consider the supergravity approximation of string theory on the warped BTZ background. Using the covariant formulation of charges we compute the background energies and $U(1)$ charge, and show that these charges match the average charges of $J\bar{T}$-deformed CFTs at finite temperature. We also reproduce the Bekenstein-Hawking entropy of the warped BTZ background from the microscopic density of states of $J\bar{T}$-deformed CFTs.

%%%%%%%%%%%%%%%%%%%%%%%%%%%%%%%%%%%%%%%%%%%%%%
\subsection{Charges and thermodynamic potentials}
%%%%%%%%%%%%%%%%%%%%%%%%%%%%%%%%%%%%%%%%%%%%%%

We will compute the zero mode charges of the warped BTZ background directly from six dimensional supergravity using the covariant formulation of charges~\cite{Barnich:2001jy,Barnich:2007bf}. Given an asymptotic Killing vector $\xi$ generating a symmetry of the theory, the infinitesimal variation of the corresponding charge is given by
  \eq{
   \d {\cal Q}_{\xi} = \frac{1}{2!\,4!} \frac{1}{8 \pi G_6} \int_{\p\ss} \ve_{\a\b\g\s\mu\nu}  \,K_{\xi}^{\mu\nu} dx^{\a}\we dx^{\b} \we dx^{\g} \we dx^{\s} , \label{variation}
  }
where $\p \ss$ is the boundary of a codimension-1 spacelike surface. The two-form $K_{\xi}^{\mu\nu}$ reads
  \eq{
  K_{\xi}^{\mu\nu} = k_{\xi,g}^{\mu\nu} + k_{\xi,B}^{\mu\nu} + k_{\xi,\Phi}^{\mu\nu},
  }
where $k_{\xi,f}^{\mu\nu}$ denotes the contribution of the field $f = \{g, B, \Phi\}$ to the variation of the charge. For the action~\eqref{sugra} the $k^{\mu\nu}_{\xi,f}$ two-forms are given by~\cite{Barnich:2001jy,Compere:2007vx}
  \eq{
  \begin{split}
  k_{\xi,g}^{\mu\nu} &= \frac{1}{2} \Big \{  \xi^{\nu}\nabla^{\mu} \d g^{\a}{}_{\a} - \xi^{\nu} \nabla_{\a} \d g^{\a\mu} + \xi_{\a} \nabla^{\nu} \d g^{\a\mu} + \frac{1}{2} \d g^{\a}{}_{\a} \nabla^{\nu} \xi^{\mu} - \frac{1}{2} \d g^{\nu\a} \nabla_{\a}\xi^{\mu} \\
  & \hspace{1cm} + \frac{1}{2} \d g^{\nu\a} \nabla^{\mu} \xi_{\a} - (\mu \lra \nu) \Big \},
  \end{split} \label{kgdef}\\
    \begin{split}
    k_{\z,B}^{\mu\nu}  & = \frac{1}{4} e^{-2\Phi} \Big \{  \big (2 H^{\mu\nu \l} \d \Phi  - \d H^{\mu\nu\l} \big ) \xi^{\a} B_{\a \l} - H^{\mu\nu\l} \xi^{\a} \d B_{\a \l} - \xi^{\mu} H^{\nu \a \l} \d B_{\a \l} 	\\
  &\hspace{2cm}  +\Big( 2\d g^{\mu \s} H_{\s}{}^{\nu\l} +  \d g^{\l\s}H^{\mu\nu}{}_{\s} - \frac{1}{2} \d g^{\s}{}_{\s} H^{\mu\nu \l}  \Big) \xi^{\a} B_{\a \l}    \\
  &\hspace{2cm}  - \d B^{\mu \l} g^{\nu \s}  \L_{\xi} B_{\s \l}  - (\mu \lra \nu) \Big \}, 
\end{split}\label{kbdef}\\
k^{\mu\nu}_{\xi,\Phi} &= \big( \xi^{\nu}\nabla^{\mu} \Phi - \xi^{\mu}\nabla^{\nu} \Phi \big) \d \Phi.
}
where $\d \equiv \d T_u \p_{T_u} + \d T_v \p_{T_v} +\d \l \p_{\l}$ and $\d f^{\a_1 \dots \a_n} = g^{\a_1 \l_1} \dots g^{\a_n \l_n} \d f_{\l_1 \dots \l_n}$ for any $f_{\a_1 \dots \a_n}$. Since the dilaton is constant the $k_{\xi,\Phi}^{\mu\nu}$ term does not contribute to the charges.

The gravitational charges depend crucially on the global properties of the background. For the warped BTZ black string with the TsT compactification of coordinates~\eqref{tstidentificationv} and \eqref{tstcirclev}, the variation of the charge reads
\eq{
   \d \Q_{\xi} = \frac{\ell^4}{8 \pi G_6} \int_{\p\ss} d\vp' d\psi d\t d\phi K_{\xi}^{t' r}  , \label{variationtst}
  }
where $\vp' = \frac{1}{2}(u + t^-)$ and $t' = \frac{1}{2}(u - t^-)$. In particular, the charges associated with the $\p_{u}$, $-\p_{t^-}$, and $\p_{\psi}$ are integrable in the space of solutions with varying $T_u$, $T_v$ and $\lambda$, and are given by
  \eq{
  \Q_{\p_{u}} =  \frac{c_{st}}{6}  T_u^2, \qquad  \Q_{-\p_{t^-}}  = \frac{c_{st}}{6} T_v^2 , \qquad  \Q_{\p_{\psi}}=0. \label{bulkcharges}
  }

The warped BTZ background has a horizon at $r_h = 2 T_u T_v$ with Bekenstein-Hawking entropy 
  \eq{
  S_{\textrm{WBTZ}}= \frac{\textrm{Area}}{4 G_6} = \frac{\pi c_{st}}{3} \big( T_u + T_v \big). \label{wbtzentropy}
  }
 The latter is independent of the deformation parameter and matches the entropy of the undeformed BTZ black string~\eqref{btzmetric}. The numerical agreement of the entropies is to be expected as the density of states of $J\bar{T}$-deformed CFTs is unchanged by the deformation. However, the fact that the background charges~\eqref{bulkcharges} are also independent of the deformation does not seem to be consistent with the spectrum of $J\bar{T}$-deformed CFTs given in eqs.~\eqref{sugawara} --~\eqref{qft_spectrumQ}.  The resolution to this puzzle lies on the fact that the warped BTZ black string is a thermal background with nonvanishing $U(1)$ charge and conjugate chemical potential.
 
As discussed in Section~\ref{se:worldsheet}, we identify the deformed $U(1)$ charge with the Killing vector $\xi_{U(1)}$ in eq.~\eqref{ju1} which, in terms of the TsT coordinates $t^-$ and $\psi$, reads
  \eq{
  \xi_{U(1)} = 2 \p_{\psi} - \l \p_{t^-}.\label{chiralU1}
  }
 One can check that this Killing vector has unit norm, $\xi_{U(1)} \cdot \xi_{U(1)} =\ell^2$, and that the corresponding gravitational charge is given by
   \eq{
   \Q_{{U(1)}} \equiv \Q_{\xi_{U(1)}}= \frac{c}{6}  \l T_v^2. \label{U1background}
   }
This charge is integrable for varying $T_u$ and $T_v$ but fixed $\l$. Since the background $U(1)$ charge is not zero, the warped BTZ black string must feature a nonvanishing chemical potential $\nu$ conjugate to $\Q_{U(1)}$. In order to determine the chemical potential, we first note that smoothness of the horizon in the Euclidean continuation of the geometry requires a thermal circle 
  \eq{
\textrm{thermal circle:} \quad (u,t^-,\psi) \sim (u + {i \pi/ T_u}, t^--{i \pi/ T_v}, \psi-2\pi i\lambda T_v),
  }
the latter of which is associated with the horizon generator 
  \eq{
  2\pi i\xi_{\textrm{th}}& = \pi i \bigg( \frac{1}{T_u} \p_u - \frac{1}{T_v} \p_{t^-} - 2 \l T_v \p_{\psi} \bigg). \label{horizongen0}
  }
We then observe that, in terms of the generator of the deformed $U(1)$ charge~\eqref{chiralU1}, the horizon generator~\eqref{horizongen0} can be written as  
%We then need to express eq.~\eqref{horizongen0} in terms of the generator of the deformed $U(1)$ charge~\eqref{chiralU1}. The horizon generator can then be written as
%To make connection to the dual field theory, it is important to identify the generator of the chiral $\xi_{U(1)}$ generator, as was argued in \eqref{suse:matching}.  As a result, the horizon generator \eqref{horizongen0} can be written in terms of the Killing vectors $\p_u,\p_{t^-}$ and $\xi_{U(1)}$ as
%
  \eq{
   2\pi i\xi_{\textrm{th}} & = \pi i \bigg [ \frac{1}{T_u} \p_u  - \bigg(\frac{1 + \l^2 T_v^2}{T_v} \bigg) \p_{t^-} - \l T_v   \xi_{U(1)}   \bigg], \label{horizongen1}
  \\
  &=
 \pi i \bigg [ \frac{1}{T_u} \big( \p_u-T_u \nu_L \xi_{U(1)} \big) - \bigg(\frac{1 + \l^2 T_v^2}{T_v} \bigg) \p_{t^-} - \big(\l T_v - \nu_L \big)  \xi_{U(1)}   \bigg] \label{horizongen}, }
where, in the second line, we introduced (by adding and subtracting) a spectral flow parameter $\nu_L$ whose value will be determined shortly. We can read the thermodynamic potentials from the coefficients of the generators $\xi_L\equiv \p_u-T_u \nu_L \xi_{U(1)}$, $\xi_R\equiv \p_{t^-}$, and $\xi_{U(1)}$, which are respectively given by 
 \eq{
  \ell {T}_L =  \frac{T_u}{\pi}, \qquad  \ell  { T}_R = \frac{1}{\pi} \frac{T_v}{1 + \l^2 T_v^2}, \qquad {\nu}=  \pi (\nu_L - \l T_v). \label{bulkpotentials}
 }
 On can easily check that these thermodynamic potentials satisfy the first law of thermodynamics~\cite{Iyer:1994ys}
  \eq{
  \delta S_{\textrm{WBTZ}}\equiv \delta \Q_{\xi_{th}}={1\over  \ell  T_L }\delta \Q_L+{1\over  \ell  T_R }\delta \Q_R + \nu \, \delta \Q_{U(1)},
  }
where the conjugate charges correspond to  % the conjugate charges are 
the gravitational charges given by $ \delta \Q_L\equiv \delta \Q_{\xi_L} $, $\delta \Q_R\equiv  \delta \Q_{\p_{t^-}}$, and the variation of eq.~\eqref{U1background}.
%and comparison with eq.~\eqref{horizongen} yields the deformed values of the temperatures and the chemical potential
  %

  %
We conclude by noting that $T_R$ does not correspond to the right-moving potential $T_v/\pi$ determined from the horizon generator in eq.~\eqref{horizongen0}. In particular, this means that the surface gravity is not proportional to the Harmonic mean of $T_L$ and $T_R$. The reason for this is that there is a mixing between the two generators $\p_{t^-}$ and $\p_\psi$ in eq.~\eqref{chiralU1} and, as a result, the would-be right-moving temperature $T_v/\pi$ is conjugate to a linear combination of $\Q_R$ and $\Q_{U(1)}$. Writing the horizon generator in terms of $\xi_{U(1)}$ as in eq.~\eqref{horizongen1} allows us to disentangle the thermodynamic potentials conjugate to these charges. We also note that the choice of spectral flow parameter $\nu_L$ does not affect the temperatures $T_{L}$ and $T_{R}$ or the $U(1)$ charge $\Q_{U(1)}$, a fact that motivates us to consider the mixed thermal ensemble discussed in Section~\ref{suse:thermodynamics}. Finally, note that $\nu_L$ cannot be arbitrary as it is constrained to yield an integrable left-moving charge, namely
  \eq{
  \delta \Q_L=\delta \Big(\frac{c_{st}}{6}T_u^2 \Big) -T_u \nu_L\delta (\lambda T_v^2).  \label{bulkqL}
  }
  %
  
%%%%%%%%%%%%%%%%%%%%%%%%%%%%%%%%%%%%%%%%%%%%%%

%%%%%%%%%%%%%%%%%%%%%%%%%%%%%%%%%%%%%%%%%%%%%%
\subsection{Matching the entropy to $J\bar{T}$} \label{suse:entropy}
%%%%%%%%%%%%%%%%%%%%%%%%%%%%%%%%%%%%%%%%%%%%%%

We now show that the entropy and charges of the deformed background match the corresponding quantities in $J\bar{T}$-deformed CFTs at finite temperature. Motivated by the bulk analysis, it is natural  to consider the mixed ensemble introduced in Section~\ref{suse:thermodynamics}. Therefore, we set the field theory temperatures $T_{L}(\l)$ and $T_{R}(\l)$ as well as the $U(1)$ charge $Q(\l)$ to the background values given in eqs.~\eqref{U1background} and~\eqref{bulkpotentials}, namely
    \eq{
 \ell {T}_L(\l) = \frac{T_u}{\pi}, \qquad \ell{T}_R(\l) = \frac{1}{\pi} \frac{T_v}{1 + \l^2 T_v^2}, \qquad Q(\l) =\Q_{U(1)}= \frac{c_{st}}{6} \l T_v^2. \label{boundarypotentials}
  }

Let us first consider the entropy in the mixed ensemble. As described earlier, the long string sector of string theory on AdS$_3$ with NS-NS flux is conjectured to be dual to a symmetric product CFT~\eqref{symprod}. In the mixed ensemble, the entropy of each copy of the seed theory ${\cal M}_{6k}$ after the $J\bar{T}$ deformation is given by~\eqref{entropyfieldtheory}
 \eq{
 {S}_{J\bar{T}}^{(i)} (T_L, T_R, Q_i) &= \frac{\pi^2 R c}{3}  \Bigg [ T_L + \frac{T_R }{\sqrt{1 - \frac{2 \pi^2  c k}{3} \mu^2 T_R^2\,}}  \Big(1 - \frac{\mu}{R}Q_i \Big) \Bigg ], \label{entropyM}
  %&=   \frac{\pi^2 R c}{3}  \Big \{ T_L(\mu) + T_R(\mu)  \Big[1 - \frac{\mu}{R}Q(0) \Big]\Big \}, 
 }
where $c = 6k$. The total entropy of the dual $J\bar{T}$-deformed CFT is obtained by summing over the contributions from each copy of the symmetric product
  \eq{
  S_{J\bar{T}} = \sum_{i = 1}^p S_{J\bar{T}}^{(i)} (T_L, T_R, Q_i). \label{totalentropy}
  }
In analogy with~\cite{Giveon:2017nie}, let us assume that for large enough temperatures, each copy in $\big ( {\cal M}_{6k} \big )^p/S_p$ contributes an equal amount to the total entropy, namely that
  \eq{
  Q_i = \frac{Q}{p} = \frac{c}{6} \l T_v^2,
  }
where $Q = \sum_i Q_i$ is the total $U(1)$ charge. Then, using the relationship between the bulk and field theory parameters given in eqs.~\eqref{mulambda} and~\eqref{dictionary}, we find that the entropy~\eqref{totalentropy} is independent of $\l$ and matches the Bekenstein-Hawking entropy of the black string
\eq{
  S_{J\bar{T}} = \frac{\pi c_{st}}{3} \big( T_u + T_v \big)=S_{\textrm{WBTZ}}.
  }

In the mixed ensemble, the average right-moving energy $E_{R}(\l)$ of the dual $J\bar{T}$-deformed CFT can be obtained from eq.~\eqref{averageER} via
 \eq{ 
 \ell E_R (\l) = p \ell E_{R,i} (\l) = \frac{c_{st}}{6} T_v^2 = \Q_{R},
 }
 where the energy $E_{R,i} (\l)$ is also assumed to be the same in each copy of the product $\big ( {\cal M}_{6k} \big )^p/S_p$. We see that the deformed right-moving energy is independent of the deformation parameter and matches the corresponding bulk charge~\eqref{bulkcharges}. Furthermore, from eq.~\eqref{averagenu} we learn that the deformed chemical potential is given by
 \eq{ 
 \nu(\l)& = -\pi \l T_v \bigg( 1 + \frac{T_v}{2T_u} \bigg), \label{nuboundary}
 }
which can be used to determine the spectral flow parameter $\nu_L$ in the bulk, namely
  \eq{
   \nu_L=- \frac{\l T_v^2}{2T_u}. \label{nuL}
  }
It is then not difficult to show that this value of $\nu_L$ makes the background left-moving charge~\eqref{bulkqL} integrable and equal to the field theory result obtained using eq.~\eqref{averageEL},
  \eq{ 
   \ell E_L(\l) = p \ell E_{L,i} (\l) =  \frac{c_{st}}{6} \bigg( T_u^2 + \frac{ \l^2 T_v^4 }{4} \bigg) = \Q_{L},
  }
where the last equality is obtained by integrating eq.~\eqref{bulkqL}. Note that our definition of the chemical potential includes an additional factor of the inverse temperature such that $\nu = \b \til{\nu}$ and the standard expression for the first law reads $\d S  = \b (\d E + \Omega\, \d J + \til{\nu} \d \Q_{U(1)})$ where $E$ and $J$ denote the energy and angular momentum while $\b$ and $\Omega$ correspond to the inverse temperature and angular potential
  \eq{
  \b = \frac{1}{2}\Big ( \frac{1}{T_L} + \frac{1}{T_R}\Big), \qquad \Omega = \frac{1}{2\b} \Big ( \frac{1}{T_L} - \frac{1}{T_R}\Big).
  }
In particular, the divergences in eqs.~\eqref{nuboundary} and~\eqref{nuL} as $T_u \to 0$ come from the divergence in the inverse temperature $\b$ such that $\til{\nu}$ remains well defined in this limit.

%
%%
%%%
\begin{comment}
Let us first consider the entropy in the mixed ensemble. Using the relationship between the bulk and field theory parameters given in eqs.~\eqref{mulambda} and~\eqref{dictionary}, we find that the entropy~\eqref{entropyfieldtheory} is independent of $\l$ and matches the Bekenstein-Hawking entropy of the warped BTZ black string~\eqref{wbtzentropy},  
  %
\eq{
  S_{J\bar{T}} = 2 \pi k \big( T_u + T_v \big)=S_{\textrm{WBTZ}}.
  }
  %
In the mixed ensemble, the average right-moving energy $E_{R}(\l)$ of $J\bar{T}$-deformed CFTs is given in eq.~\eqref{averageER} and satisfies
 %
 \eq{ 
 \ell E_R (\l) = k T_v^2 = \Q_{R},
 }
 %
which is independent of the deformation parameter and matches the corresponding bulk charge~\eqref{bulkcharges}. Furthermore, from eq.~\eqref{averagenu} we learn that the deformed chemical potential is given by
 %
 \eq{ 
 \nu(\l)& = -\pi \l T_v \bigg( 1 + \frac{T_v}{2T_u} \bigg),
 }
 %
which can be used to determine the spectral flow parameter $\nu_L$ in the bulk, namely
  %
  \eq{
   \nu_L=- \frac{\l T_v^2}{2T_u}. \label{nuL}
  }
  %
It is then not difficult to show that this value of $\nu_L$ makes the background left-moving charge~\eqref{bulkqL} integrable and equal to the field theory result,
  %
  \eq{ 
   \ell E_L(\l) = k \bigg( T_u^2 + \frac{ \l^2 T_v^4 }{4} \bigg) = \Q_{L},
  }
  %
where the first equality follows from the field theory analysis~\eqref{averageEL} while the second one is obtained from integrating the bulk charge~\eqref{bulkqL}.
\end{comment}
%%%
%%
%

We can show that the charges computed above satisfy the $J\bar{T}$ spectrum given in eqs.~\eqref{sugawara} --~\eqref{qft_spectrumQ}. In order to accomplish this we need to compute the charges before the deformation. From the relationship between the deformed and undeformed temperatures and chemical potential~\eqref{deformedTL} --~\eqref{deformednu}, we find 
  \eq{
  \ell T_L(0)={T_u\over \pi}, \qquad \ell T_R(0)={T_v\over\pi},\qquad Q(0) =-\frac{c_{st}}{6}\lambda T_v^2.
  }
According to the discussion in Section~\ref{se:worldsheet}, the undeformed $U(1)$ charge $Q(0)$ is related to the bulk Killing vector $2\p_\chi$. Indeed, the gravitational charge corresponding to $2\p_\chi$ matches the field theory value of $Q(0)$, namely
   \eq{
   \Q_{2\p_\chi} = -\frac{c_{st}}{6} \lambda T_v^2=Q(0).
   }
Note that the $\Q_{2\p_\chi}$ charge is integrable provided that $\d \l=0$. Finally, using the standard thermodynamic relations between the temperatures and energies of the undeformed theory~\eqref{cftthermo} we obtain
\eq{
E_L(0) &=E_L(\lambda), \qquad E_R(0)=E_R(\lambda), \qquad Q(0) = - Q(\l), \label{wbtzelrq}
}
formulae that are compatible with the spectrum of $J\bar T$-deformed CFTs. 

It may seem confusing to have an undeformed $U(1)$ charge that depends on the deformation parameter. However, recall that both $\Q_{2\p_{\chi}}$ and $\Q_{U(1)}$ are integrable only for fixed $\l$ and we should not interpret eq.~\eqref{wbtzelrq} as general flow equations with varying $\l$. Instead, for a given value of $\l$, eq.~\eqref{wbtzelrq} tells us that the energies and $U(1)$ charge of the warped BTZ background are related to states in the undeformed theory satisfying
  \eq{
  Q(0) = -\l \ell E_R(0).
  }
  %
%In other words, the undeformed ``image'' of the warped BTZ background is not the undeformed BTZ black string, even though these backgrounds are related by a TsT transformation.
This is consistent with property $(v)$ of the $J\bar{T}$ spectrum discussed in Section~\ref{suse:fieldtheoryspectrum}.

In conclusion, by carefully identifying the deformed $U(1)$ charge and the thermodynamic potentials, we found that the Bekenstein-Hawking entropy of the warped BTZ background with the TsT compactification of coordinates matches the microscopic entropy of $J\bar{T}$-deformed CFTs. Furthermore, the matching of the gravitational charges with the thermodynamics induced by the $J\bar{T}$ deformation provides nontrivial evidence that the warped BTZ background is dual to a thermal state in the dual $J\bar{T}$-deformed CFT.

%%%%%%%%%%%%%%%%%%%%%%%%%%%%%%%%%%%%%%%%%%%%%%

%%%%%%%%%%%%%%%%%%%%%%%%%%%%%%%%%%%%%%%%%%%%%%
\section{Warping global AdS} \label{se:globalads}
%%%%%%%%%%%%%%%%%%%%%%%%%%%%%%%%%%%%%%%%%%%%%%

In this section we describe the TsT transformation of global AdS$_3 \times S^3$. In analogy to the finite-temperature case, we show that this background is locally equivalent to a marginal deformation of the worldsheet, write down the worldsheet spectrum, and compute the supergravity charges. We note that these expressions can be obtained from analytic continuation of the results derived in Sections~\ref{se:warpedbtz},~\ref{se:worldsheet}, and~\ref{se:entropy}.

%%%%%%%%%%%%%%%%%%%%%%%%%%%%%%%%%%%%%%%%%%%%%%
\subsection*{The warped AdS$_3$ background}
%%%%%%%%%%%%%%%%%%%%%%%%%%%%%%%%%%%%%%%%%%%%%%

In our gauge the global AdS$_3\times S^3$ metric is given by\footnote{The AdS$_3$ factor can be expressed in a more standard gauge such as $ds_3^2= \ell^2 \big(  d\rho^2  - \cosh^2\rho \,dt^2  + \sinh^2\rho\, d\vp^2 \big)$ via the change of coordinates $r = \sinh^2\rho + \frac{1}{2}$, $u = \vp + t$, and $v = \vp - t$.}
  \eq{
 {ds^2}{} &=  \ell^2 \bigg \{ \frac{dr^2}{ 4r^2 - 1} + r du dv - \frac{1}{4} du^2 -\frac{1}{4} dv^2 + d\Omega_3^2  \bigg \}, \qquad r \ge \frac{1}{2}, \label{globalmetric}
  }
where $(u,v) \sim (u + 2\pi, v + 2\pi)$ and the background $B$-field and dilaton are given in eqs.~\eqref{mbtzbfield} and~\eqref{mbtzdilaton}. The metric~\eqref{globalmetric} can be obtained from the BTZ black hole~\eqref{btzmetric} after analytic continuation of the temperatures, 
  \eq{
  T_u \to -\frac{i}{2}, \qquad T_v \to \frac{i}{2}. \label{globaltemp}
  }
Eq.~\eqref{globaltemp} plays an important role in this section as the results reported herein can be derived from Sections~\ref{se:warpedbtz},~\ref{se:worldsheet}, and~\ref{se:entropy} by analytic continuation of the temperatures. 

We now perform the same TsTs transformation described in Sections~\ref{se:jtbar} and~\ref{se:warpedbtz}. Namely, we T-dualize along $\chi$, shift $v \to t^- + \frac{2 \l}{k} \psi$ and $\chi \to \psi$, T-dualize again along $\psi$ and finally shift the coordinates by
  \eq{
  t^- = v + \frac{\l}{2} \chi, \qquad \psi = \chi - \frac{\l}{4} v.  \label{shift33}
  }
The resulting warped AdS$_3$ background is given by  
  \eq{
  \begin{split}
 ds^2_{st} \!&=  \ell^2 \bigg \{ {dr^2\over 4 r^2 - 1} + \frac{1}{2} \big( 2r du - dv \big) \big[ dv + \b \big( d \chi + \cos\t\, d\phi \big) \big] \!+ \frac{1}{4} dv^2 \!- \frac{1}{4} du^2 \! + d\Omega_3^2  \bigg \}, \label{globalmetricdef} \end{split} \\
 B & = {\ell^2 \over4} \Big \{ \cos \t\, d\phi \we d \chi - \big( 2 r du - dv \big) \we \big[ dv +  \b \big( d \chi + \cos\t\,d\phi \big) \big ] - \frac{\l}{2} dv \we d\chi \Big \}, \label{globalbfielddef}\\
e^{- 2\Phi} &= 1 - \frac{\l^2}{4},\label{globaldilatondef}
  }
where the parameter $\b$ is the analytic continuation of $\a$ in eq.~\eqref{alphadef} and satisfies
\eq{
\b =  \frac{4\l}{4 - \l^2}.
}

Finally, we note that the worldsheet action on the warped AdS$_3$ background described by eqs.~\eqref{globalmetricdef} and~\eqref{globalbfielddef} can be obtained from eq.~\eqref{btzactiondef1} by analytic continuation of the temperatures. In particular, the deformed worldsheet action $S_{\l}$ satisfies the same differential equation given in eq.~\eqref{diffdef1}, namely
   \eq{
   \frac{\p S_{\l} }{\p \l} =  \frac{1}{k} \int d^2z\,  (j_{\p_{t^{-}}})^a  (j_{{\xi_{U(1)}}})^b \e_{ab}, \label{diffdefglobal}
   }
where the Noether currents $j_{\p_{t^-}}$ and $j_{\xi_{U(1)}}$ can be obtained from analytic continuation of eqs.~\eqref{jtm} and~\eqref{ju1}. 
 
%%%%%%%%%%%%%%%%%%%%%%%%%%%%%%%%%%%%%%%%%%%%%%

%%%%%%%%%%%%%%%%%%%%%%%%%%%%%%%%%%%%%%%%%%%%%%
\subsection*{The deformed string spectrum}
%%%%%%%%%%%%%%%%%%%%%%%%%%%%%%%%%%%%%%%%%%%%%%

In analogy to Section~\ref{suse:wbtzspectrum}, the spectrum of strings on the warped AdS background~\eqref{globalmetricdef} can be derived using spectral flow. The crucial step in the derivation is the observation that the deformed theory is equivalent to an undeformed one satisfying twisted boundary conditions, a relationship made manifest through a nonlocal change of coordinates. The latter is given by eqs.~\eqref{vhat1} and~\eqref{chihat2} after analytic continuation. Omitting the intermediate steps, the zero modes of the stress tensor are found to satisfy
  \eq{
  \hat{L}_0 = &\til{L}_0 +  p_w w_u + \frac{k}{4} w_u^2  + q_w w_{\chi} - k w_{\chi}^2  +  \frac{\b^2}{1 + \b^2} \frac{1}{k} \bigg (\bar{p}_w^2 - \frac{{q}_w^2}{4}  + \frac{  \bar{p}_w {q}_w}{\b} \bigg),   \label{globalL0} \\
  \hat{\bar{L}}_0  = & \til{\bar{L}}_0 + \bar{p}_w w_v + \frac{k}{4} w_v^2 +   \frac{\b^2}{1 + \b^2} \frac{1}{k} \bigg (\bar{p}_w^2 - \frac{{q}_w^2}{4}  + \frac{  \bar{p}_w {q}_w}{\b} \bigg) ,\label{globalL0bar}
  }
where $\til{L}_0$ and $\til{\bar{L}}_0$ denote the zero modes before the deformation, and the winding momenta $p_w$, $\bar{p}_w$, and $q_w$ are defined as the zero modes of the Noether currents $j_{\p_u}$, $j_{\p_v}$, and $j_{\p_\chi}$ modulo topological $\t_{(x^a)}$ terms. The derivation of $\hat{L}_0$ and $\hat{\bar{L}}_0$ is valid for any representation of $SL(2,R)$, including its discrete (real $j$) and continuous (complex $j$) representations. Not surprisingly, eqs.~\eqref{globalL0} and~\eqref{globalL0bar} can be obtained from the expressions derived in the warped BTZ background by analytic continuation of the temperatures. We also note that these equations match the spectrum of classical winding strings on the warped AdS background where $\til{L}_0$ and $\til{\bar{L}}_0$ depend on the parameters of the solution.

Equations~\eqref{globalL0} and~\eqref{globalL0bar} have not yet taken into account the global properties of the warped AdS spacetime and are valid for arbitrary compactifications of coordinates. We are particularly interested in the TsT compactification. For a string winding along the spatial circle~\eqref{tstcirclev} the winding parameters satisfy
  \eq{
  w_u = w, \qquad w_v = \frac{4w}{ 4 + \l^2} , \qquad w_{\chi} = \frac{ \l w}{ 4 + \l^2}, \qquad w \in \mathbb{Z}. \label{globaltstwinding}
  }
This implies that, in terms of the winding momenta, the physical energies conjugate to translations along $u$ and $t^-$, as well as the undeformed $U(1)$ charge, satisfy
  \eq{
  \ell E_L &  = p_w + \frac{k w}{4}, \qquad \ell E_R = \frac{1}{4 + \l^2} \Big[4 \bar{p}_w - \l  q_w + k w \Big],\qquad q= q_w - \frac{2k \l w}{4 + \l^2}.\label{q2}
  }
Imposing the Virasoro constraints before and after the deformation yields the same warped spectrum found in the zero and finite-temperature backgrounds given in eqs.~\eqref{ogspectrumleft} --~\eqref{ogspectrumu1}. This is not surprising given that these equations are independent of the temperatures. In particular, this means that we can reproduce the spectrum of $J\bar{T}$-deformed CFTs when $w = -1$, $\ell \l = \mu k$, and $\ell = R$. 

\subsection*{Matching the background charges to $J\bar T$}
Finally, we propose the following map between the field theory charges of the $J\bar{T}$-deformed theory and the gravitational charges of the warped AdS background  
  \eq{
  \ell E_L(\lambda)-{Q(\lambda)^2\over 4 k} &= \Q_{\p_{u}} = -\frac{c_{st}}{24}, \\
    \ell E_R(\lambda) &=\Q_{-\p_{t^-}}  = -\frac{c_{st}}{24}, \\
    Q(\lambda) &= \Q_{{U(1)}} = -\frac{c_{st}}{24} \l. \label{adsbulkcharges}
  }
In particular, we note that the undeformed and deformed charges are related, as in the finite temperature case, by eq.~\eqref{wbtzelrq}. In analogy with the discussion in Section~\ref{se:entropy}, the charges of the warped AdS background can be interpreted as the result of deforming a special state in a theory with a fixed value of $\l$.
 
%%%%%%%%%%%%%%%%%%%%%%%%%%%%%%%%%%%%%%%%%%%%%%

%%%%%%%%%%%%%%%%%%%%%%%%%%%%%%%%%%%%%%%%%%%%%%
%%%%%%%%%%%%%%%%%%%%%%%%%%%%%%%%%%%%%%%%%%%%%%
%%%%%%%%%%%%%%%%%%%%%%%%%%%%%%%%%%%%%%%%%%%%%%
 
 \bigskip

%\begin{comment}
%%%%%%%%%%%%%%%%%%%%%%%%%%%%%%%%%%%%%%%%%%%%%%
\section*{Acknowledgments}
%%%%%%%%%%%%%%%%%%%%%%%%%%%%%%%%%%%%%%%%%%%%%%
We are grateful to Soumangsu Chakraborty, Pankaj Chaturvedi, Gaston Giribet, Amit Giveon, Monica Guica, David Kutasov, Mukund Rangamani, Eva Silverstein, and Boyang Yu for helpful discussions. We thank the Institut de Physique Th\'eorique at CEA Saclay for hospitality during the development of this project as well as the Tsinghua Sanya International Mathematics Forum for hospitality during the workshop ``Black holes and holography''. We thank the Simons Center for Geometry and Physics for providing a stimulating environment during the workshop ``$T\bar{T}$ and Other Solvable Deformations of Quantum Field Theories''. We are also grateful to the Yukawa Institute for Theoretical Physics at Kyoto University  where this work was completed for hospitality during the workshop YITP-T-19-03 ``Quantum Information and String Theory 2019". This work was supported by the National Thousand-Young-Talents Program of China and NFSC Grant No.~11735001. The work of L. \!A. was also supported by the International Postdoc Program at Tsinghua University.
%%%%%%%%%%%%%%%%%%%%%%%%%%%%%%%%%%%%%%%%%%%%%%
%\end{comment}

%%%%%%%%%%%%%%%%%%%%%%%%%%%%%%%%%%%%%%%%%%%%%%
%%%%%%%%%%%%%%%%%%%%%%%%%%%%%%%%%%%%%%%%%%%%%%
%%%%%%%%%%%%%%%%%%%%%%%%%%%%%%%%%%%%%%%%%%%%%%

\bigskip

%%%%%%%%%%%%%%%%%%%%%%%%%%%%%%%%%%%%%%%%%%%%%%
\appendix
%%%%%%%%%%%%%%%%%%%%%%%%%%%%%%%%%%%%%%%%%%%%%%

%%%%%%%%%%%%%%%%%%%%%%%%%%%%%%%%%%%%%%%%%%%%%%
\section{Geodesics and winding strings in warped BTZ} \label{se:classicalstrings}
%%%%%%%%%%%%%%%%%%%%%%%%%%%%%%%%%%%%%%%%%%%%%%
 
In this appendix we derive the spectrum of classical point particles and winding strings on the warped BTZ background. We will see that the contributions of the deformation are nonlocal and match the worldsheet derivation of the spectrum given in Section~\ref{se:worldsheet}. Our analysis is also valid for the warped AdS background after the analytic continuation~\eqref{globaltemp}.

%%%%%%%%%%%%%%%%%%%%%%%%%%%%%%%%%%%%%%%%%%%%%%
\subsection*{Point particle spectrum} 
%%%%%%%%%%%%%%%%%%%%%%%%%%%%%%%%%%%%%%%%%%%%%%
 
Let us consider a geodesic $x^{\mu} \equiv x^{\mu}(\tau)$ in the warped BTZ background with dimensionless momenta $p$, $\bar{p}$, and $q$ along the $u$, $v$, and $\chi$ coordinates such that
  \eq{ 
   { p} = g_{u \s} \dot{x}^{\s}, \qquad  \bar{p} =  -g_{v \s} \dot{x}^{\s}, \qquad  \frac{{q}}{2} =  g_{\chi \s} \dot{x}^{\s}.
  }
The geodesic satisfies
  \eq{
  \dot{u} &= -\frac{1}{k}  \frac{ 2 \big( \bar{p} r + 2 {p} T_v^2 \big)}{r^2 - 4T_u^2 T_v^2}, \label{ugeod}\\
  \dot{v} &= -\frac{1}{k}  \frac{1}{1-\a^2}\frac{1}{T_v^2} \big ( \bar{p} + \a  {q}T_v \big ) + \frac{1}{k} \frac{r}{r^2 - 4T_u^2 T_v^2}\frac{1}{T_v^2} \big ( \bar{p} r + 2 {p} T_v^2 \big ), \label{vgeod} \\
   \dot{\chi} &= \frac{1}{k}  \frac{2}{1 - \a^2} \bigg ( {q} + \frac{\a}{T_v} \bar{p} \bigg ), \label{chigeod}
  }
where we assumed that $\t = \frac{\pi}{2}$ and $\phi = 0$, while $r$ obeys
  \eq{
  k^2 \ddot{r} - \frac{ k^2 r \dot{r}^2 - 16 r \big(  \bar{p}^2 T_u^2 + p^2 T_v^2 \big )}{r^2 - 4T_u^2 T_v^2} + 8 p \bar{p} \bigg(\frac{r^2 + 4T_u^2 T_v^2}{r^2 - 4T_u^2 T_v^2}\bigg) = 0. \label{req}
  }
The generic solutions to eq.~\eqref{req} are independent of $\l$ and are given by
  \eqst{
  r_{\pm} = \frac{1}{2 c_1 k ^2 } \Big \{ {-8 p\bar{p}} +  e^{\pm \sqrt{c_1}(\tau + c_2)} + 4 (4 {p}^2 - c_1 k^2   T_u^2 \big ) \big( 4 \bar{p}^2 - c_1 k^2 T_v^2 \big) e^{\mp \sqrt{c_1} ( \tau + c_2)} \Big \},
  }
where $c_1$ and $c_2$ are integration constants. In particular, we note that $c_1$ is proportional to the conformal weight in the undeformed worldsheet theory and its sign depends on the nature of the geodesic, being strictly positive (negative) for spacelike (timelike) geodesics. On the other hand the constant $c_2$ is fixed by boundary conditions.

The worldsheet stress tensor of the deformed theory can be easily read from the metric. Its left-moving component is given by
  \eqsp{
  T = - k \bigg \{ &\frac{ (\p r )^2}{4(r^2 - 4T_u^2 T_v^2)} + ( r \p u + 2 T_v^2 \p v ) \bigg [ \p v + \frac{\a}{2 T_v} ( \p \chi + \cos\t\,\p\phi ) \bigg] - T_v^2 ( \p v )^2  \\
  & + T_u^2 ( \p u )^2 + \frac{1}{4} \Big [ (\p \chi + 2 \cos\t\,\p\phi ) \p \chi + (\p \t )^2 + (\p \phi )^2 \Big] + \dots, \label{leftstresstensor}
  }
where we have omitted the contributions of the ${\cal M}^4$ and the right-moving component $\bar{T}$ is obtained by letting $\p \to \bp$.\footnote{Recall that the worldsheet deformation is marginal. Hence, while the global symmetries of the background are changed, the worldsheet theory is still conformal. In particular, the left and right-moving components of the worldsheet stress-energy tensor $T(z)$ and $\bar{T}(\bar{z})$ are chirally conserved.} The zero modes of the stress tensor $L_0 = - \frac{1}{2\pi} \oint T$ and $\bar{L}_0 = - \frac{1}{2\pi} \oint \bar{T}$ satisfy
  \eq{
  L_0 &= \til{L}_0 + \frac{\a^2}{1 - \a^2} \frac{1}{k} \bigg [ \frac{\bar{p}^2}{4 T_v^2} + \frac{{q}^2}{4}  + \frac{ \bar{p} {q}}{2 \a T_v} \bigg], \label{L0geodesic}  \\
  \bar{L}_0  &= \til{\bar{L}}_0 + \frac{\a^2}{1 - \a^2} \frac{1}{k} \bigg [ \frac{\bar{p}^2}{4 T_v^2} + \frac{{q}^2}{4}  + \frac{ \bar{p} {q}}{2 \a T_v} \bigg], \label{Lbar0geodesic}
  }
where $\til{L}_0$ and $\til{\bar{L}}_0$ are independent of $\a$ and given by
  \eq{
  \til{L}_0 =  \frac{c_1 k}{16} + \frac{{q}^2}{4 k} + \til{L}_0^{{\cal M}^4}, \qquad \til{\bar{L}}_0= \frac{c_1 k}{16} + \frac{{q}^2}{4 k} + \til{\bar{L}}_0^{{\cal M}^4}, \label{undefL0til}
  }
with $\til{L}_0^{{\cal M}^4}$ and $\til{\bar{L}}_0^{{\cal M}^4}$ denoting the contributions of the internal ${\cal M}^4$ coordinates.

Imposing the classical Virasoro constraints $L_0 = \bar{L}_0 = 0$ yields the spectrum of point particles with nonvanishing momenta $p$, $\bar{p}$, and $q$ in the deformed background. Note that the presence of the right moving momentum $\bar{p}$ in eqs.~\eqref{L0geodesic} and~\eqref{Lbar0geodesic} implies that the deformed theory is nonlocal along the right moving coordinate $v$, as expected from the $J\bar{T}$ deformation of the dual CFT~\cite{Guica:2017lia}. Finally note that the $\a$-dependent contribution is independent of $c_1$ and, up to conventions, is consistent with the result found in~\cite{Azeyanagi:2012zd}.

%%%%%%%%%%%%%%%%%%%%%%%%%%%%%%%%%%%%%%%%%%%%%%

%%%%%%%%%%%%%%%%%%%%%%%%%%%%%%%%%%%%%%%%%%%%%%
\subsection*{Winding string spectrum} 
%%%%%%%%%%%%%%%%%%%%%%%%%%%%%%%%%%%%%%%%%%%%%%
 
We now generate new solutions to the worldsheet equations of motion by ``winding up'' the geodesics obtained in the previous section. Given any solution $x^{\mu}(z,\bar{z})$ to the equations of motion we can obtain new solutions $x_w^{\mu}(z,\bar{z})$ via 
  \eq{
  u_w &= u + w_u z \label{windingsolution1} \\
  v_w &= v - \frac{w_v}{1 - \a^2}\big( \bar{z} + \a^2 {z} \big ) - \frac{2\a w_{\chi}}{ T_v(1 - \a^2)} \tau, \label{windingsolution2} \\
    \qquad \chi_w &= \chi + \frac{2 w_{\chi}}{1 - \a^2} \big( {z} + \a^2 \bar{z} \big) + \frac{4 \a T_v w_v}{1 - \a^2} \tau, \label{windingsolution3}  \\
    r_w& = r, \qquad \qquad \t_w = \t, \qquad \qquad \phi_w = \phi, \label{windingsolution4}
  }
where $w_u$, $w_v$, and $w_{\chi}$ are constants independent of the worldsheet coordinates. The novel shifts in eqs.~\eqref{windingsolution2} and~\eqref{windingsolution3} are consequences of the mixing between the $v$ and $\chi$ coordinates induced by the deformation of the worldsheet. The new solutions $x^{\mu}_w$ can be interpreted as strings winding along the $u$, $v$ and $\chi$ coordinates such that
  \eq{
 \oint \p_{\s} u_w = 2\pi w_u, \qquad \oint \p_{\s} v_w = 2\pi w_v , \qquad   \oint \p_{\s} \chi_w = 4\pi w_{\chi}, \label{windingbc}
  }
where we have assumed $\oint \p_\s x^{\mu} = 0$. Note that the nature of the winding parameters $w_u$, $w_v$, and $w_\chi$ depends on the global properties of the deformed background. In what follows we derive the general form of the spectrum valid for any values of these parameters, a result that can be applied to backgrounds satisfying different compactifications of coordinates.

Classical winding string solutions were also obtained from geodesics in global AdS$_3$ in ref.~\cite{Maldacena:2000hw}. Therein, the winding solution $x^{\mu}_w$ --- to which eqs.~\eqref{windingsolution1} --~\eqref{windingsolution4} reduce to in the $\a \to 0$ and $w_\chi \to 0$ limits --- was linked to spectral flow of the affine $SL(2,R)_L \times \widebar{SL(2,R)}_R$ symmetry algebra. Spectral flow is an essential ingredient of string theory in AdS$_3$ generating altogether new representations of $SL(2,R)$. As described in Section~\ref{se:worldsheet}, spectral flow also plays an important role in the warped BTZ background since the contributions to the spectrum arising from the worldsheet deformation are equivalent to spectral flow in the undeformed theory. 

In terms of the winding momenta defined in eq.~\eqref{zeromodecharges}, the zero modes of the stress-energy tensor satisfy
  \eq{
 {L}_0 = &\til{L}_0  +  p_w  w_u - k ( T_u  w_u )^2  + q_w w_{\chi}  - k w_{\chi}^2  +   \frac{\a^2}{1 - \a^2} \frac{1}{k} \bigg ( \frac{\bar{p}_w^2}{4 T_v^2} + \frac{{q}_w^2}{4}  + \frac{  \bar{p}_w {q}_w}{2 \a T_v} \bigg),   \label{L0geodesic2}  \\
   {\bar{L}}_0  = & \til{\bar{L}}_0  + \bar{p}_w w_v - k ( T_v  w_v)^2 +    \frac{\a^2}{1 - \a^2} \frac{1}{k} \bigg ( \frac{\bar{p}_w^2}{4 T_v^2} + \frac{{q}_w^2}{4}  + \frac{  \bar{p}_w {q}_w}{2 \a T_v} \bigg), \label{Lbar0geodesic2}
  }
where the undeformed zero modes $\til{L}_0$ and $\til{\bar{L}}_0$ are given in eq.~\eqref{undefL0til}. Setting $L_0 = \bar{L}_0 = 0$ leads to the spectrum of classical winding strings with momenta along the $u$, $v$, and $\chi$ coordinates. The $\a$-dependent contributions to the spectrum remain nonlocal and are independent of the undeformed values $\til{L}_0$ and $\til{\bar{L}}_0$. Finally note that eqs.~\eqref{L0geodesic2} and~\eqref{Lbar0geodesic2} match the expressions derived from the worldsheet in eqs.~\eqref{L0} and~\eqref{L0bar}.

%%%%%%%%%%%%%%%%%%%%%%%%%%%%%%%%%%%%%%%%%%%%%%

%%%%%%%%%%%%%%%%%%%%%%%%%%%%%%%%%%%%%%%%%%%%%%
\section{Chiral current-current deformation} \label{se:currentcurrent}
%%%%%%%%%%%%%%%%%%%%%%%%%%%%%%%%%%%%%%%%%%%%%%

In this appendix we consider a background that is closely related to the warped BTZ black string described in eqs.~\eqref{btzmetricdef} and~\eqref{btzbfielddef}. This background is also obtained from a TsT transformation of the BTZ black string but it is accompanied by the change of coordinates~\eqref{shift22} and, crucially, an additional \emph{locally} gauge transformation
  \eq{
   B \to B - \frac{\ell^2}{2} \l T_v^2 dv \we d \psi. \label{diffgauge}
  }
The resulting spacetime has been previously studied in~\cite{Detournay:2012dz,Azeyanagi:2012zd} and is only locally equivalent to the warped BTZ black string considered in this paper. This follows from the fact that the would-be gauge transformation~\eqref{diffgauge} is not compatible with the TsT identification of coordinates in eqs.~\eqref{tstidentificationv} and~\eqref{tstcirclev}. The distinction is important as the global properties of the warped BTZ background play an important role in the derivation of the spectrum of winding strings and its thermodynamics. In particular, it is only on the TsT background satisfying the TsT compactification of coordinates where we can reproduce the spectrum and thermodynamics of $J\bar{T}$-deformed CFTs. 

We now note that the background~\eqref{btzmetricdef} and~\eqref{btzbfielddef} with~\eqref{diffgauge} can be interpreted as a marginal deformation of the worldsheet by the product of two chiral currents. Indeed, in terms of the $\bar{j}^1$ and $k^3$ currents, which remain chirally conserved in the deformed theory, we find that the action is given by
  \eq{
  S'_{\l} = S'_0 + \dd S'_\l,
  }
where the undeformed action $S'_0$ reads
   \eqsp{
  S_{0}' = k \int dz^2 \bigg\{ & \frac{\p r \bar{\p} r}{4 \big(r^2 - 4 T_u^2 T_v^2 \big)} +  r \p v \bp u  +  T_v^2 \p v \bp v  + T_u^2 \p u \bar{\p} u + \frac{1}{4} \p\t\bp\t + \frac{1}{4} \p\phi\bp\phi \\
  &+\frac{1}{4} \big(\p\chi + 2 \cos\t\,\p\phi \big) \bp \chi \bigg \},
  }
while the deformation $ \dd S'_\l$ satisfies
  \eq{
  \dd S_{\l}' = \frac{2 \a}{k} \int d^2z\,   \bar{j}^1(\bar{z}) k^3(z). \label{jtbar2}
  }
The $\bar{j}^1$ and $k^3$ currents are given in eqs.~\eqref{k30} and~\eqref{j10} and are independent of the deformation. Hence, the deformed action $S'_{\l}$ also satisfies a differential equation after an appropriate normalization of the chiral currents. Note that the worldsheet deformation~\eqref{jtbar2} is exactly marginal~\cite{Chaudhuri:1988qb} and reduces to its zero temperature counterpart~\eqref{jtbar} in the $T_v \to 0$ limit. Furthermore, at linear order in $\lambda$, eq.~\eqref{jtbar2} is given by
  \eq{
  \dd S_{\l}' \to \frac{2 \l}{k} \int d^2z\, [2T_v \bar{j}^1(\bar{z})] k^3(z) , \label{jtbar3}
  }
and takes the same asymptotic form as the marginal deformation that corresponds to a single-trace version of the $J\bar{T}$ operator (see Section~\ref{se:jtbar}). These observations indicate further connections between the warped BTZ background and the single-trace $J\bar{T}$ deformation of the dual CFT.

%%%%%%%%%%%%%%%%%%%%%%%%%%%%%%%%%%%%%%%%%%%%%%

%%%%%%%%%%%%%%%%%%%%%%%%%%%%%%%%%%%%%%%%%%%%%%
% Bibliography
%%%%%%%%%%%%%%%%%%%%%%%%%%%%%%%%%%%%%%%%%%%%%%

\ifprstyle
	\bibliographystyle{apsrev4-1}
\else
	\bibliographystyle{JHEP}
\fi

\bibliography{jtbar}

%%%%%%%%%%%%%%%%%%%%%%%%%%%%%%%%%%%%%%%%%%%%%%

%%%%%%%%%%%%%%%%%%%%%%%%%%%%%%%%%%%%%%%%%%%%%%
% The end
%%%%%%%%%%%%%%%%%%%%%%%%%%%%%%%%%%%%%%%%%%%%%%

\end{document}

%%%%%%%%%%%%%%%%%%%%%%%%%%%%%%%%%%%%%%%%%%%%%%
%%%%%%%%%%%%%%%%%%%%%%%%%%%%%%%%%%%%%%%%%%%%%%
%%%%%%%%%%%%%%%%%%%%%%%%%%%%%%%%%%%%%%%%%%%%%%